\documentclass[reprint,amsmath,amssymb,aps,prl,longbibliography]{revtex4-1}

\usepackage{color}
\usepackage{graphicx}% Include figure files
\usepackage{dcolumn}% Align table columns on decimal point
\usepackage{bm}% bold math
\usepackage{braket}

\usepackage{units}
\usepackage[caption=false]{subfig}

\let\Re\relax
\let\Im\relax
\DeclareMathOperator{\Re}{Re}
\DeclareMathOperator{\Im}{Im}

\renewcommand{\vec}[1]{\mathbf{#1}}

\DeclareMathOperator{\sgn}{sgn}

\begin{document}

\title{Persistent Hall response in a quantum quench}

\author{Justin H.\ Wilson$^{1}$}
\email{jwilson@caltech.edu}
\author{Justin C. W. Song$^{1,2}$}
\author{Gil Refael$^{1,2}$}
\affiliation{$^1$ Institute of Quantum Information and Matter and Department of Physics,} 
\affiliation{$^2$ Walter Burke Institute of Theoretical Physics, California Institute of Technology, Pasadena, CA 91125 USA}

\date{\today}

\begin{abstract}
Out-of-equilibrium systems can host phenomena that transcend the usual restrictions of equilibrium systems.
Here we unveil how out-of-equilibrium states, prepared via a quantum quench, can exhibit a non-zero Hall-type response that persists at long times, and even when the instantaneous Hamiltonian is time reversal symmetric; both these features starkly contrast with equilibrium Hall currents. 
Interestingly, the persistent Hall effect arises from processes beyond those captured by linear response, and is a signature of the novel dynamics in out-of-equilibrium systems.
We propose quenches in two-band Dirac systems as natural venues to realize persistent Hall currents, which exist when {\it either} mirror or time-reversal symmetry are broken (before or after the quench).
Its long time persistence, as well as sensitivity to symmetry breaking, allow it to be used as a sensitive diagnostic of the complex out-equilibrium dynamics readily controlled and probed in cold-atomic optical lattice experiments. 
\end{abstract}

\pacs{05.70.Ln, 67.85.-d, 73.43.-f}

\maketitle

The subtle quantum coherence encoded in the topology of crystal wavefunctions is responsible for a wide array of robust quantum phenomena~\cite{VonKlitzing1986,Nagaosa2010,Stormer1999,Klitzing1980}, e.g. the quantum Hall effect.  
While these concepts originated in the solid-state, cold atoms have recently become a system of choice for experimentally unraveling topology on the microscopic level~\cite{Lin2011,Bloch2012,Will2015} due to the array of new probes available. 
For example, these probes have been used to image the skipping orbits (edge-states) in a cold-atomic quantum Hall system \cite{Stuhl2015}, directly measure the Berry curvature~\cite{Jotzu2014}, and Zak phase~\cite{Atala2013} in cold-atomic topological bands. 

\begin{figure}
  \includegraphics[width=\columnwidth]{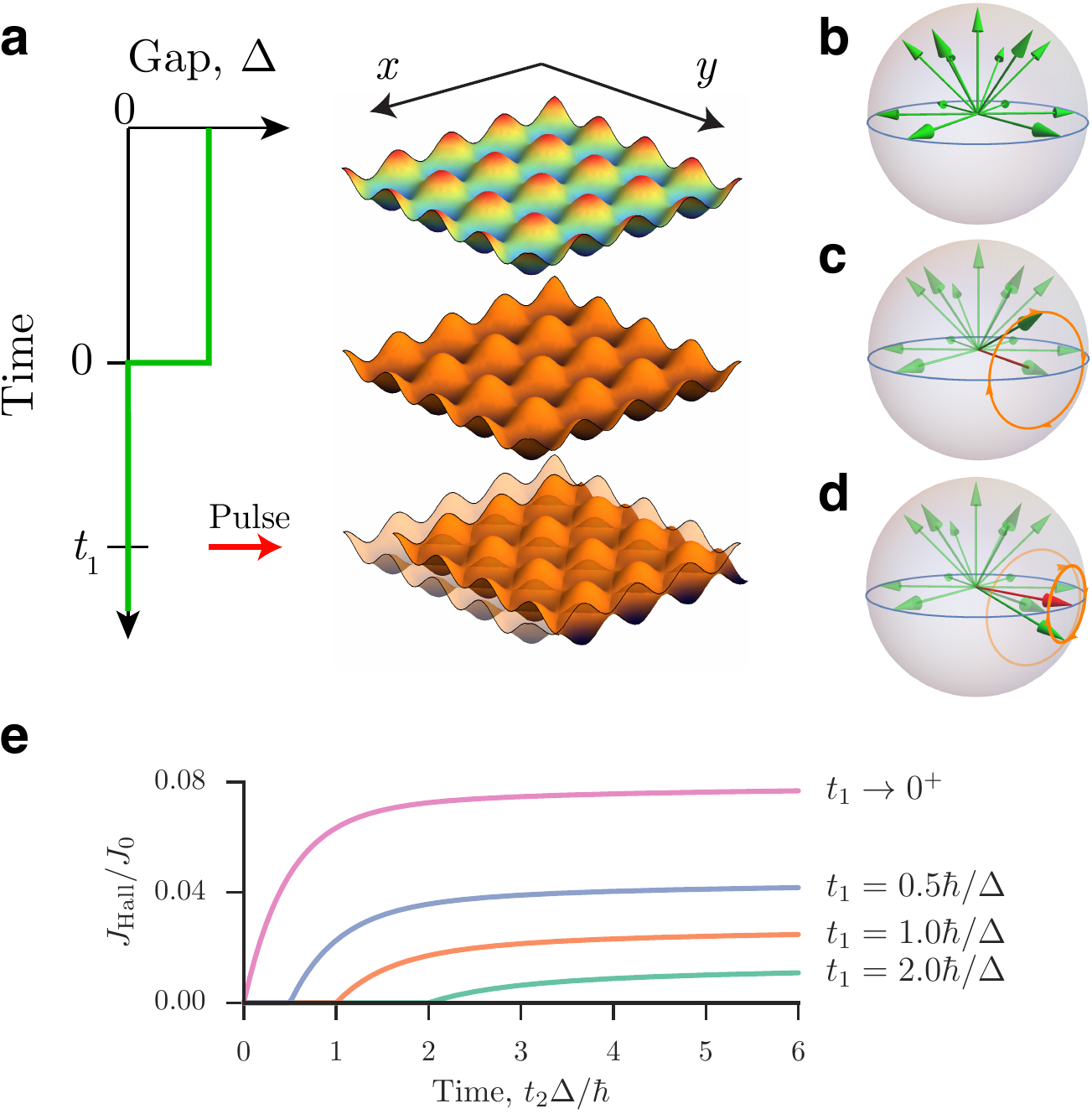}
  \caption{{\bf a.} Quantum quenches implemented in cold-atomic optical lattices, where a parameter in the Hamiltonian is changed suddenly shown by change in color of optical lattice. 
  {\bf b.} Pseudospinors on a Bloch sphere prepared in the Haldane state {\bf c}. exhibit Larmor precession after the Hamiltonian is quenched into zero gap. 
  {\bf d.} The pseudospinors can acquire a transverse shift after the system is pulsed in the longitudinal direction. 
  {\bf e.} Persistent Hall response (orange) for quenching protocol described in Eq.~\ref{eq:haldane} (and panel {\bf b-d}). 
  Here, the green curve shows $\Delta (t)$ quench and characteristic $J_0 = \frac{e^2}{h} \frac{\Delta^2}{e \hbar v_{\mathrm F}}$.}
 \label{fig:protocol}
\end{figure}

One readily available tool is the \emph{quantum quench}. 
A state is prepared in the many-body ground state of a Hamiltonian $H(\zeta)$. 
After which, a physical parameter $\zeta$ (e.g.\ lattice depth, detuning) is changed {\it suddenly} (Fig. \ref{fig:protocol}a), setting the system into dynamical evolution far from equilibrium \cite{Greiner2002}. 
The ease with which distinct Hamiltonians can be accessed via quenches and driving in general opens up a tantalizing possibility of achieving new out-of-equilibrium phenomena with no equilibrium analog \cite{Oka2009,Kitagawa2011,DAlessio2014,Rudner2014,FoaTorres2014,Budich2015,Caio2015,Dehghani2015}.

Here we unveil a completely new type of dynamical response achieved in out-of-equilibrium states (OES) which can be prepared via quantum quenches. 
In particular, we show that certain OES can feature an unconventional Hall current even when the instantaneous Hamiltonian preserves time-reversal symmetry (TRS). 
Intriguingly, when a short-time pulsed electric field is applied to OES, the Hall current generated persists even long after the pulse application, saturating to a non-zero value at long times (Fig.~\ref{fig:protocol}). 
These characteristics have no analog in equilibrium systems, and, as we argue, originate from coherent evolution of the wavefunction after a quantum quench. 

The origin of the unconventional, quench-induced response can be most easily illustrated for non-interacting and clean Dirac systems, where many-body states can be represented as a collection of pseudospinors on a Bloch sphere (Fig.~\ref{fig:protocol}b-d). 
In these, a state is prepared in the ground state of a Dirac Hamiltonian $H(\Delta)$, with TRS breaking gap $\Delta$ (Fig.~\ref{fig:protocol}a).  
At $t=0$, the Hamiltonian is quenched to $H(\Delta = 0)$ [where TRS is preserved], yielding dynamics for OES, with the pseudospinors exhibiting Larmor precession (Fig.~\ref{fig:protocol}c). 

To probe OES, a {\it short} pulse of strength $\mathbf A = \int dt \, \mathbf E(t)$ can be applied to the system at time $t=t_1$ (Fig.~\ref{fig:protocol}{\bf a},{\bf e}), shifting the Larmor orbits along $\vec E$. 
Averaged over one-cycle, longitudinal momentum along $\vec E$ increases. 
However, in addition to this, the constraint of pseudospinors being on the Bloch sphere allows a {\it transverse} shift to accumulate. 
As a result, at long times $t=t_2$, we obtain an unconventional Hall current
\begin{align}
  \mathbf J_{\mathrm{Hall}}(t_1, t_2\rightarrow\infty) =  \Sigma_{\rm Hall}^\infty(t_1) \hat{\mathbf z} \times\mathbf A,
  \label{eq:hall}
\end{align}
that persists long after the pulse $\mathbf E(t)$ as shown in Fig.~\ref{fig:protocol}e. 
Here $\Sigma_{\rm Hall}^\infty$ is non-universal function depending on $t_1$ and model specifics described below. 
Additionally, while we use the language of electromagnetic response, in cold-atom optical lattices $e \mathbf A$ can be easily effected by a shift in momentum $\Delta \mathbf p$ brought on by a sudden force; in such systems $\mathbf J_{\mathrm{Hall}}$ takes the form of a particle current.

Hall currents from OES (Eq.~\eqref{eq:hall}, Fig.~\ref{fig:protocol}{\bf e}) are strikingly different to those found in equilibrium systems. 
Hall currents generated by short pulses in the latter vanish at long times after the pulse is applied, are dissipationless, and do not involve the shift of the Fermi sea. 
In contrast, OES Hall currents persist even at long times, and involve overall momentum shifts of the entire Fermi sea. 
As a result, we expect that when the Fermi sea relaxes, the Hall current will degrade. 

Further, we consider the non-interacting and clean (disorder-free) limit to highlight the role coherent evolution of the wavefunction has in forming $\mathbf J_{\rm Hall}$ (see Larmor precession in Fig.~\ref{fig:protocol}{\bf b}-{\bf d}). 
A useful analogy with the coherent evolution of spins in nuclear magnetic resonance (NMR) protocols can be drawn, where decay of the NMR signal can be used as a sensitive diagnostic of scattering, for e.g. spin-spin, spin-environment relaxation. 
In the same way, we anticipate that the decay profile of $\vec J_{\rm Hall}$ that arises from coherent pseudospinor evolution can be used as a diagnostic of relaxation and/or thermalization processes in OES when interactions are allowed. 

The ease with which Dirac-type \cite{Jotzu2014} and other spin-orbit coupled Hamiltonians \cite{Lin2011} can be constructed in setups for ultra-cold bosons and fermions allows these effects to be easily accessed---though we find that fermions more readily see these effects. 
In order to observe the Hall effect and separate it from an overwhelming longitudinal response, we propose a time-of-flight setup in the direction perpendicular to the applied pulse while keeping a confining potential in the direction of the applied pulse.
In such an experimental set-up, the gap, as tuned by Zeeman coupling or ``shaking'' of the cold atom lattice, is suddenly turned off.
The ``pulse'' is then implemented some time after the quench by applying a sudden and brief force upon the system (e.g.\ tilting the confining potential for a very short time).

Let us now explain the effect with a two-band Hamiltonian $H(\Delta) = \sum_\mathbf p c^\dagger_{\mathbf p} h(\mathbf p, \Delta) c_{\mathbf p}$ with
\begin{align}
  h(\mathbf{p},\Delta) = \epsilon_0(\mathbf{p}) \mathbb{I} + 
  \mathbf d(\mathbf p,\Delta(t)) \cdot \bm \sigma,
  \label{eq:hamiltonian}
\end{align}
where $\mathbf p = (p_x,p_y)$ is the two-dimensional momentum and $\bm \sigma = (\sigma_x,\sigma_y,\sigma_z)$ are the Pauli matrices, and $\Delta (t)$ is a gap parameter that varies as a function of time. 
When $\mathbf d(\mathbf p,\Delta(t))$ varies slowly, the equilibrium wavefunction describes the properties of $H$. 
However, when $\mathbf d(\mathbf p,\Delta(t))$ changes rapidly as in a quantum quench, the response depends intimately on the evolution of the wavefunction. 

To illustrate this, we first analyze a simple example that captures the essential physics - a quenched, single-cone, low-energy Haldane-type model - obeying Eq.~\eqref{eq:hamiltonian} with
\begin{align}
\epsilon_0(\mathbf{p}) = 0, \quad \mathbf d(\mathbf p,\Delta(t)) = (p_x, p_y,\Delta \Theta[-t]),
\label{eq:haldane}
\end{align}
where $\Theta(t)$ is the Heaviside function.
This captures the essential physics of the usual two-cone Haldane model, hence the name. 
The physics described below does not change if we use a two-cone Haldane model, the only difference being an extra degeneracy factor of two.
If, on the other hand, a gap was created by breaking inversion symmetry, the effects described here would be identically zero.
When $t<0$, it is convenient to describe the system [Eqs.~\eqref{eq:hamiltonian},\eqref{eq:haldane}] by the familiar ground state wavefunction of the Haldane model in a single cone
$
\ket{\Psi_0} = \prod_{\mathbf p} \left( \cos (\theta_{\mathbf p}/2) \ket{+} - e^{i\phi_{\mathbf p}} \sin(\theta_{\mathbf p}/2) \ket{-}\right),
$
where $\cos\theta_{\mathbf p} = -\Delta/\sqrt{p^2 + \Delta^2}$ and $p e^{i\phi_{\mathbf p}} = p_x + i p_y$. 
This wavefunction is characterized by a Chern number $\mathcal C = \int\frac{d^2 p}{(2\pi)^2} \hat{\mathbf{z}}\cdot \nabla_{\vec p} \times \braket{ u_{\vec p}| i \nabla_{\vec p} | u_{\vec p}}$.
For the half-filled band in $\ket{\Psi_0}$ we have $\mathcal{C} = 1/2$ per flavor. 
As a result, $\ket{\Psi_0}$ features a bulk Hall conductivity of $\sigma_{xy} = \mathcal{C} e^2/h$.
When $t>0$, the wavefunction evolves with $\ket{\Psi_1(t)} = \mathcal{U}(t)\ket{\Psi_0}$, with $\mathcal{U}(t) = {\rm exp}[-i H  t] $, so that
\begin{equation}
\ket{\Psi_1(t)} = \prod_{\mathbf p} \ket{\psi_1(\mathbf p)}, \;  \ket{\psi_1(\mathbf p)}= f_t\ket{+} - e^{i\phi_{\mathbf p}} g_t \ket{-}
\label{eq:wavefunction2}
\end{equation}
where
$ f_t  = \cos\tfrac{\theta_{\mathbf p}}2 \cos p t + i \sin\tfrac{\theta_{\mathbf p}}2 \sin p t,$ and    $g_t   = \sin\tfrac{\theta_{\mathbf p}} 2\cos p t + i \cos\tfrac{\theta_{\mathbf p}}2 \sin p t$.
Interestingly, at every point in time, the instantaneous wavefunction, $\ket{\Psi_1(t)} $, is still characterized by the same Chern number $\mathcal C = 1/2$ as $\ket{\Psi_0}$ before the quench \cite{DAlessio2014,Caio2015}. 
However, as we argue below, the current response becomes disconnected from $\mathcal{C}$ [i.e. the equilibrium bulk current responses corresponding to $\mathcal C$ described above no longer apply]. 
Instead, $\ket{\Psi_1}$ is characterized by an unconventional current response.

To extract the response properties of Eq.~\eqref{eq:wavefunction2} we consider the following pulse-type protocol [see Fig.~\ref{fig:protocol}] where (i) at $t=t_1$ a short pulse [$E_x(t) = A_x \delta(t-t_1)$] is applied to the system so that $\mathbf p \rightarrow \mathbf p - e\mathbf A$, (ii) and the Hall current, $\vec J_{\rm Hall}$, that develops is measured at $t=t_2$. Here $t_1,t_2 >0$ occur after the quench. 
We note that after the pulse at $t_1$, the Hamiltonian in Eq.~\eqref{eq:haldane} changes $\mathbf d(\mathbf p,0)  \rightarrow \mathbf d(\mathbf p - e \mathbf A, 0)$. 
As a result, the wavefunction in Eq.~\eqref{eq:wavefunction2} continues to evolve as $\ket{\Psi_2 (t_2)} = \prod_{\vec p} \ket{\psi_2(\mathbf p)}$, with $\ket{\psi_2(\mathbf p)} = e^{-i(t_2-t_1) h(\mathbf p-e\mathbf A,0)} \ket{\psi_1(\mathbf p,t_1)}$.

The current response can be obtained via $\vec J = \braket{\Psi| \hat{\vec j} |\Psi}$, where $\hat{\vec j} = \partial H/\partial \vec A$. 
Using $\ket{\Psi} = \ket{\Psi_2(t_2) }$, Eq.~\eqref{eq:haldane},\eqref{eq:wavefunction2}, and extracting the component of $\vec J$ transverse to the applied field $\vec E$, we obtain $\vec J_{\rm Hall}$ as shown in Fig.~\ref{fig:protocol}e. 
Here, $\vec J_{\rm Hall}$ was obtained via numerical integration with a pre-quench $\ket{\Psi_0}$ where the entire valence band was filled. 
A full discussion of $\vec J$ is contained in the supplement~\cite{supplement}. Due to the collective action of all electrons in the valence band, $\mathbf J_{\rm Hall}$ does not have an apparent oscillatory structure in Fig.~\ref{fig:protocol}e.

Strikingly, $\vec J_{\rm Hall}$ in Fig.~\ref{fig:protocol}e grows from zero (when the pulse is first applied at $t_1$) and saturates at long times to a non-vanishing value, $\vec J_{\rm Hall} (t_1,t_2\to \infty) = \vec J_{\rm Hall}^\infty (t_1)$ as seen in Fig.~\ref{fig:protocol}e. 
As we argue below, this behavior is generic for OES and its qualitative behavior is independent of model specifics.
The non-zero $\vec J_{\rm Hall}^{\infty} (t_1)$ is unconventional and arises from the near-lockstep Larmor precession of the pseudospinors $\ket{\psi_1(\vec p)}$ in Eq.~\eqref{eq:wavefunction2} that form the full many-body OES $\ket{\Psi_1}$. 

\begin{figure}
      \includegraphics[width=\columnwidth]{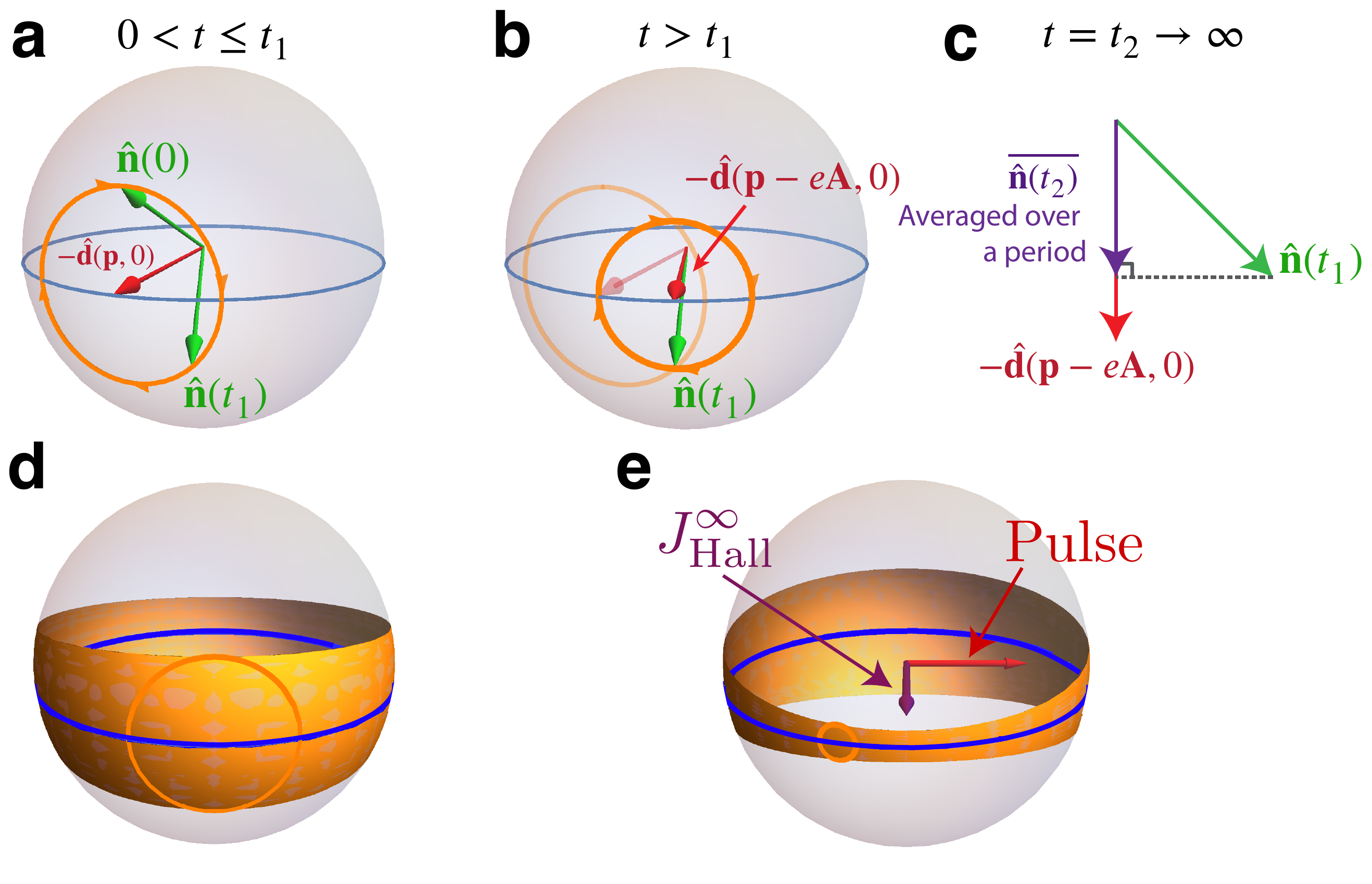}
  \caption{
  {\bf a.} After quench, the state $\hat{\mathbf{n}}$ Larmor precesses on the Bloch sphere. 
  {\bf b.} After the pulse, the center of Larmor precession shifts  due to the momentum boost $e\mathbf{A}/c$.  
  {\bf c.} For long times, the shift in the state's average over a Larmor period, $\overline{\hat{\mathbf n}(t_2)}$, persists leading to a  current at $t_2 \to \infty$. 
  {\bf d.} For the Haldane model, Eq.~\eqref{eq:haldane}, the orange manifold represents the combined Larmor orbits of states with the same $|\mathbf p|$. 
  {\bf e.} After the pulse, the manifold of Larmor orbits changes to give a perpendicular shift in the average $\overline{\hat{\mathbf n}(t_2)}$ resulting in $J^\infty_{\rm Hall}$.
    }
  \label{fig:BlochSphereExplanation}
\end{figure}

We can understand this geometrically by considering Larmor precession of the pseudospins on the Bloch sphere. 
Even though we are interested in quenches defined in Eq.~\eqref{eq:haldane}, the following geometrical analysis is general and applies to two-band models.
Mapping each spinor onto the Bloch sphere via $\hat{\mathbf n} = \braket{\psi_1(\mathbf p) | \bm \sigma |\psi_1(\mathbf p)} $, we can describe the
Larmor precession of the spinors via the equations of motion:
\begin{align}
  \partial_t \hat{\mathbf n} & = 2 \mathbf d(\mathbf p, 0) \times \hat{\mathbf n}, \quad \hat{\mathbf n}(t=0) = -\hat{\mathbf d}(\mathbf p, \Delta).
\end{align}
To understand \emph{why} this implies a persistent current, consider a ring of momenta with $|\mathbf p| = p$ held constant.
With Larmor precession for $t>0$, they will oscillate around a point on the equator, see Fig.~\ref{fig:BlochSphereExplanation}{\bf a,d}.
Then, at time $t=t_1$ we apply a pulse.
As shown by the red arrow in Fig.~\ref{fig:BlochSphereExplanation}b, the pulse has the effect of shifting the center of rotation for Larmor precession $\mathbf d(\mathbf p, 0) \rightarrow \mathbf d(\mathbf p - e\mathbf A, 0)$.
As a result, at long times the shift in average $\hat{\mathbf n}$ persists (see Fig.~\ref{fig:BlochSphereExplanation}{\bf b,c,e}). 
Since the direction of the pseudospin, $\hat{\mathbf n}$ directly corresponds to the direction of current flow in the Haldane model, a Hall current can persist at long times. 

We note that the long-time average of $\hat{\mathbf n}$ is just its projection at time $t_1$ along the new precession direction $\mathbf d(\mathbf p - e\mathbf A, 0)$ yielding $[\hat{\mathbf n}(t_1) \cdot \hat{\mathbf d}(\mathbf p - e\mathbf A, 0)]\hat{\mathbf d}(\mathbf p - e\mathbf A, 0)$.
Next we note that the current operator is $\hat j_{\mu} = -e\partial_{p_\mu} h(\mathbf p - e\mathbf A, 0) = -e \partial_{p_\mu}\mathbf d(\mathbf p - e\mathbf A, 0) \cdot \bm \sigma$. 
As a result, the projection of average $\hat{\mathbf n}$ along $\partial_{p_\mu}\mathbf d(\mathbf p - e\mathbf A, 0)$ yields the current.
Combining these, we obtain an expression for the long-time current for a single momentum state $\mathbf p$
\begin{align}
  j_\mu^\infty(\mathbf p,t_1) = -e[\hat{\mathbf n}(\mathbf p,t_1) \cdot \hat{\mathbf d}(\mathbf p - e\mathbf A, 0)] \partial_{p_\mu}d(\mathbf p - e\mathbf A, 0). \label{eq:exact-persistent-current-singlestate}
\end{align}
The expression in Eq.~\eqref{eq:exact-persistent-current-singlestate} is independent of a specific two-band model \footnote{A two-band model neglecting current contributions from $\epsilon_0(\mathbf p)$; however, those contributions do not have a Hall response.}.

We now consider the quench specified in Eq.~(\ref{eq:haldane}) so that $\mathbf n(t_1) = \braket{\psi_1|\bm \sigma |\psi_1}$ reads as $\mathbf n(t_1) = -\mathbf p \sin \theta_{\mathbf p} + (\cos 2 pt_1 \, \hat{\mathbf z} - \sin 2 p t_1\, \hat{\mathbf z} \times \hat{\mathbf p}) \cos \theta_{\mathbf p}$ (using Eq.~\eqref{eq:wavefunction2}).
Integrating over all $\mathbf p$ (for a filled band prior to quench), we obtain a total current
\begin{equation}
 J^\infty_\mu(t_1) = -e \int \frac{d^2 p}{(2\pi \hbar)^2} \frac{\hat{\mathbf n}(t_1) \cdot (\mathbf p - e \mathbf A)}{|\mathbf p - e\mathbf A|} \partial_{p_\mu} |\mathbf p - e\mathbf A|.
 \label{eq:full}
\end{equation}
While this quantity can be fully evaluated (see supplement for discussion), for brevity and to capture the essential physics, we expand Eq.~\eqref{eq:full} in $\mathbf{A}$.
Discarding terms that integrate to zero we arrive at an expression for the Hall current as
\begin{equation}
  \mathbf J^\infty_{\rm Hall} = -e^2 \int \frac{d^2 p}{(2\pi \hbar)^2} \frac{p_x^2}{p^3} \frac{\Delta}{\sqrt{p^2 + \Delta^2}} \sin 2 p t_1 (\mathbf A \times \hat{\mathbf z}).  \label{eq:Infinite-Hall}
\end{equation}
After an angular integral, we obtain Eq.~\eqref{eq:hall} with $\Sigma_{\rm Hall}^\infty(t_1) = -\frac{e^2}{2h} \frac{\Delta}{\hbar}\int_0^{\frac{\pi}2} dz\, e^{-2\frac{t_1 |\Delta|}{\hbar} \sin z} $.

While $\ket{\psi_1(\vec p)}$ with similar energies precess with frequencies that are close to each other, over long times $t_1$, small differences in their precession frequency allow their Larmor orbits to slowly drift out of phase, degrading $\vec J_{\rm Hall}^\infty (t_1)$. 
Analyzing Eq.~\eqref{eq:Infinite-Hall} for large $t_1$, we obtain
\begin{align}
    \vec J_{\rm Hall}^\infty (t_1) = -\sgn(\Delta) \frac{e^2}{4h} \frac{A_x}{t_1} + O(t_1^{-2}),
\end{align}
which shows that the longer we wait after the quench to pulse the system, the smaller $\vec J_{\rm Hall}^\infty (t_1)$, as evidenced in the diminishing $\vec J_{\rm Hall}$ current profiles shown in Fig.~\ref{fig:protocol}{\bf e}. 
This aging behavior is a characteristic of the different energies of the pseudospinors that form pre-quench $\ket{\Psi_0}$.

Importantly, persistent $\vec J_{\rm Hall}^\infty$ does not occur in equilibrium systems; in fact, it is disallowed due to the existence of a finite DC conductivity even without disorder. 
To see this, consider the response in equilibrium captured by $j_y(t) = \int \sigma_{yx}(t-t') E_x(t') dt'$. 
For a pulse $E_x (t) = A_x \delta (t)$, we have $j_y(t) = \sigma_{yx}(t) A_x$. Thus $\sigma_{yx}^{\mathrm{DC}} = \frac1{A_x}\int j_y(t) dt$.
As a result, for $\sigma_{yx}^{\mathrm{DC}}$ that is finite (e.g., the anomalous and conventional Hall effect, the quantum Hall effect), then $j_y(t)\rightarrow 0$ as $t\rightarrow \infty$ due to integrability.

OES Hall currents in Eq.~\eqref{eq:hall} depend intimately on the underlying symmetries of the Hamiltonian, $h$, in Eq.~\eqref{eq:hamiltonian}. 
In particular, we find $\Sigma_{\rm Hall}^\infty $ depends on the {\it absence} of either mirror, $M_y^{-1} h(p_x,p_y) M_y = h(p_x, -p_y)$, or time-reversal, $T^{-1} h(-\mathbf p) T = h(\mathbf p)$, symmetry. 
To expose this, we analyze the contribution of $\vec p$ states to the persistent response in Eq.~\eqref{eq:exact-persistent-current-singlestate}. Expanding in the pulse strength $\mathbf A$, we obtain $j_\mu^\infty(\mathbf p,t_1) \approx \chi^\infty_{\mu \nu}(\mathbf p,t_1) A_\nu$. Indeed $\Sigma_{\rm Hall}^\infty = \int d \vec p \chi^\infty_{\rm Hall} (\vec p)$, where $\chi^\infty_{\rm Hall} = \frac12(\chi^\infty_{yx} - \chi^\infty_{xy})$. 
Writing $\mathbf d_0 = \mathbf d(\mathbf p, 0)$ yields $\chi^\infty_{\rm Hall} = \chi_M^\infty + \chi_T^\infty $, where $\chi_M^\infty = e^2\partial_{[p_y} d_0 \partial_{p_x]} \hat{\mathbf d}_0 \cdot \hat {\mathbf d}\cos 2d_0 t_1$, and $\chi_{T}^\infty =  -e^2\partial_{[p_y} d_0 \partial_{p_x]} \hat{\mathbf d}_0 \cdot \hat{\mathbf d}_0 \times \hat{\mathbf d} \sin 2 d_0 t_1$. 
Here the brackets $\partial_{[p_y} \cdots \partial_{p_x]}$ denote antisymmetrization, and $M$ and $T$ subscripts denote contributions controlled by $M_y$ and $T$.  
Importantly, if $h$ possesses $M_y$-symmetry, then $\chi_M^\infty(p_x,p_y) = - \chi_M^\infty(p_x,-p_y)$. On the other hand, if $h$ possesses $T$-symmetry, then $\chi_T^\infty(\mathbf p) = - \chi_T^\infty(-\mathbf p)$ (see supplement~\cite{supplement}). 
As a result, when $h$ satisfies both $M_y$ and $T$ symmetries (before and after quench), opposing momentum states will give contributions of opposite sign, and $\Sigma_{\rm Hall}^\infty =  \int d \vec p \chi^\infty_{\rm Hall} (\vec p) = 0$. 
Hence, finite $\Sigma_{\rm Hall}^\infty$ arises from breaking of {\it either} $M_y$ or $T$ symmetry before or after the quench \footnote{Indeed, $h$ in Eq.~\ref{eq:haldane} possesses $M_y$ symmetry (anti-unitary symmetry), but $h(t<0)$ breaks $T$ symmetry resulting in the observed Hall current.} in contrast to the symmetry requirements for Hall currents in equilibrium linear response \cite{Sodemann2015}.

\begin{figure}
  \includegraphics[width=\columnwidth]{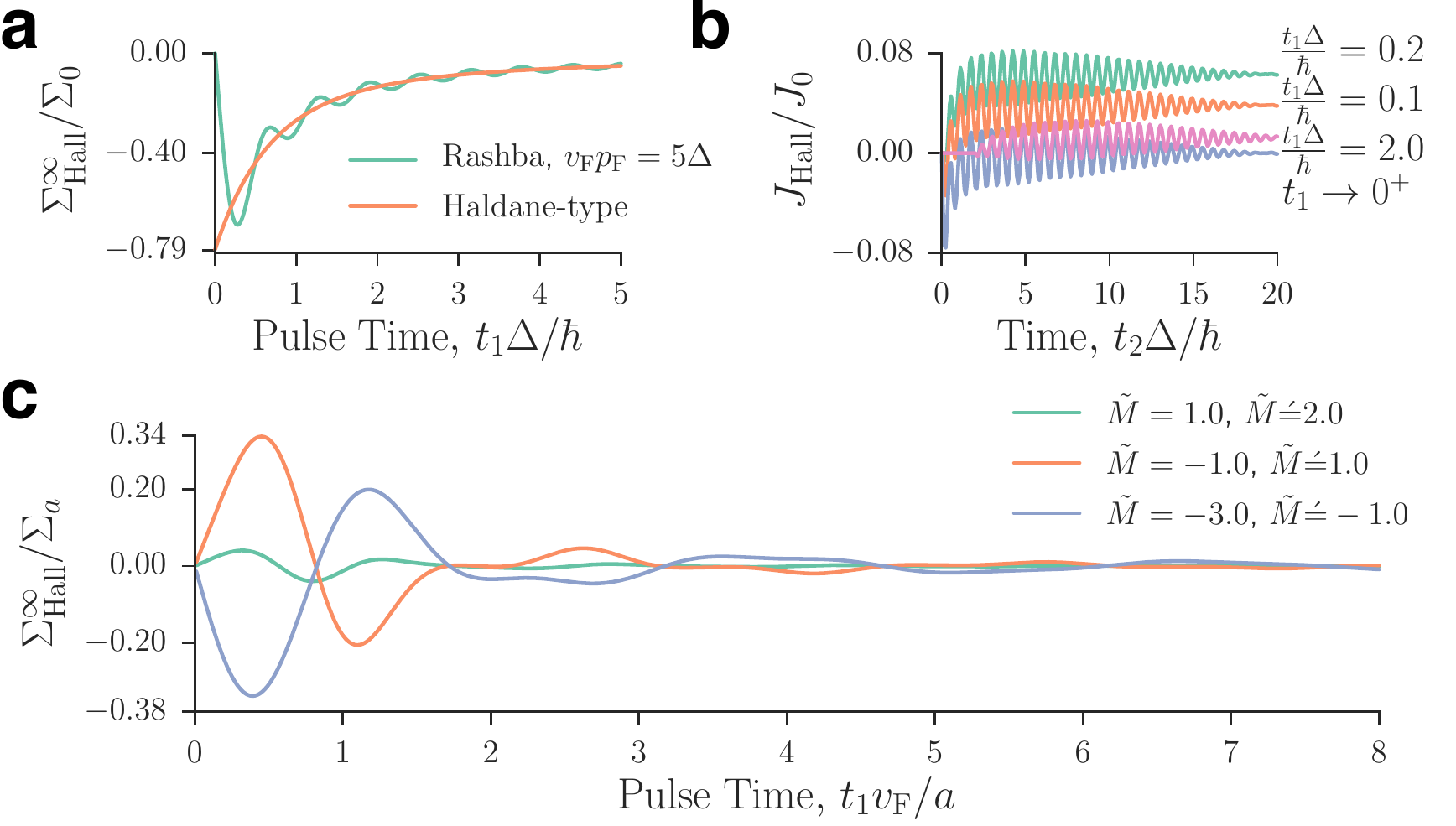}
  \caption{Other models for OES Hall current. {\bf a.} The long-time persistent $\Sigma_{\mathrm{Hall}}^\infty$  dies off as a function of pulse time $t_1$ for Haldane and Rashba model. 
  The Fermi momentum acts as a cutoff for Rashba, causing the oscillations and $\Sigma_{\rm Hall} \rightarrow 0$ as $t_1\rightarrow 0$. 
  {\bf b.} For the Rashba model, the current evolves in an oscillatory way due to the cutoff $p_{\rm F}$. 
  {\bf c.} Persistent $\Sigma_{\mathrm{Hall}}^\infty$ in the half-BHZ model (see text) sees similar oscillations due to the cutoff provided by the square lattice. 
  Interestingly $\Sigma_{\mathrm{Hall}}^\infty \neq 0$, regardless of the phase we begin or end in.
  For the Rashba model, we used $e A_x = 0.1 \Delta/v_{\rm F}$ and $v_{\rm F} p_{\rm F} = 5 \Delta$. 
  In the above, characteristic $J_0=\frac{e^2}{h} \frac{\Delta^2}{e \hbar v_{\mathrm F}}$, $\Sigma_0 = \frac{e^2}{h}\frac{\Delta}{\hbar}$, $\Sigma_a = \frac{e^2}{h} \frac{v_{\mathrm{F}}}{a}$, and $\tilde M = M \frac{a}{\hbar v_{\mathrm F}}$.\label{fig:othermodels}}
\end{figure}

While OES Hall response is disconnected from the Chern number, $\mathcal{C}$, $\Sigma_{\mathrm{Hall}}^\infty$ can still be expressed in terms of bulk band properties.
In particular, for $M_y$ symmetric Hamiltonians with a filled band prior to quench, we find an equivalent TKNN-like formula
\begin{align}
 \Sigma_{\mathrm{Hall}}^\infty = -e^2 \int \frac{d^2 p}{(2\pi)^2} \partial_{t_1} \Omega_{p_y p_x} \log d(\mathbf p, 0),\label{eq:berrycurv-persistent}
\end{align}
where $\Omega_{p_y p_x} = \frac12 \hat{\mathbf n}(t_1) \cdot (\partial_{p_y} \hat{\mathbf n}(t_1) \times \partial_{p_x}\hat{\mathbf n}(t_1))$ is the Berry curvature of the evolved $\mathbf p$ state evaluated at pulse time $t_1$. 
While arising from Berry curvature, we note that it is manifestly distinct from $\mathcal{C}$ and is not quantized.

Finally, we examine other quench protocols for Eq.~\eqref{eq:hamiltonian}. 
As we will see, these yield similar responses to the Haldane protocol examined above. 
One interesting example is a Rashba type protocol where
\begin{align}
  \epsilon_0(\mathbf p) = \frac{p^2}{2m},  \quad \mathbf d(\mathbf p, \Delta) = (-v_{\mathrm F} p_y, v_{\mathrm F}  p_x, \Delta\Theta(-t)), \label{eq:RashbaHamiltonian}
\end{align}
and chemical potential $\mu = 0$. 
As shown in Fig.~\ref{fig:othermodels}a,b, the Rashba protocol also yields a Hall current that persists at long times. 
Interestingly, the Hall current in Fig.~\ref{fig:othermodels}a exhibits an oscillatory behavior which arises from the momentum cutoff of Eq.~\eqref{eq:RashbaHamiltonian} at $p_{\rm F}=v_{\mathrm{F}}[2 m(m v_{\mathrm{F}}^2 + \sqrt{m^2 v_{\mathrm{F}}^4 + \Delta^2})]^{1/2}$; this contrasts with the smooth behavior of Fig.~\ref{fig:protocol}e, which had no momentum cutoff. 

For $t_2 \to \infty$, the Hall current response levels out (Fig.~\ref{fig:othermodels}a,b). 
Indeed, its persistent response, $\mathbf J_{\mathrm{Hall}}^\infty$, matches the Haldane protocol closely (see Fig.~\ref{fig:othermodels}a), except in one important way. 
In the Rashba protocol, it takes a finite $t_1$ for the $ \mathbf J_{\mathrm{Hall}}^\infty$ to ``turn-on'': magnitude $\mathbf J_{\mathrm{Hall}}^\infty$ increases from zero at small $t_1$, and decreases at long $t_1$. 
In contrast, the Haldane protocol featured $\mathbf J_{\mathrm{Hall}}^\infty$ that was maximal at $t_1 \to 0^+$. 
This difference also arises due to the momentum cutoff which does not appear in the low-energy model of Eq.~\eqref{eq:haldane} where there exist states on the Bloch sphere that have already performed multiple Larmor orbits even for an infinitesimal $t_1$, yielding a large $\mathbf J_{\mathrm{Hall}}^\infty$. 

Quench type protocols exhibiting $ \mathbf J_{\mathrm{Hall}}^\infty$ can also be realized in lattice models. 
In these, the bands are finite as opposed to the continuum bands discussed above. 
We illustrate such a protocol for a ``half-BHZ'' type model in a square lattice~\cite{Bernevig2013}, wherein Eq.~\eqref{eq:hamiltonian} takes   $ \epsilon_0(\mathbf p) = 0$ and $\mathbf d(\mathbf p, M(t)) = \tfrac{\hbar v_{\mathrm F}}{a}(\sin \tfrac{a p_x}{\hbar}, \sin \tfrac{ a p_y}{\hbar}, M(t) + 2 - \cos \tfrac{a p_x}{\hbar} - \cos \tfrac{a p_y}{\hbar}) $. 
Here $M(t<0) = M$ and $M(t>0) = M'$ represents the quench, and $a$ is the lattice constant. 
In the ground state, this model has different topological phases represented by $M$ \footnote{$M>0$ and $M<-4$ are trivial with equilibrium $\sigma_{xy} = 0$, $-2<M<0$ is a topological insulator with equilibrium $\sigma_{xy} = -1$, and $-4<M<-2$ is also a topological insulator with equilibrium $\sigma_{xy} = +1$.}
Picking $M,M'$ values allows to quench within and between the trivial and topological phases, yielding a persistent Hall current as well (Fig.~\ref{fig:othermodels}c).
As in the case of the Rashba Hamiltonian, there is ``turn-on'' behavior with time scale corresponding to the momentum cutoff provided by $a^{-1}$. 

The general framework, as well as the specific model realizations, presented here demonstrate that OES prepared via quench can manifest Hall currents that persist long after the application of an excitation pulse. 
Strikingly, they occur under different symmetry requirements than that found in equilibrium systems and can arise even when the instantaneous Hamiltonian is TRS preserving. 
The experimental conditions necessary for probing OES are readily available in current cold atom setups \cite{Stamper-Kurn2013}: the persistent, quench-induced Hall currents described can be measured via time-of-flight and provides a new diagnostic of coherent wavefunction dynamics.
The Hall response of OES depend intimately on the entire history of wavefunction evolution unlike in equilibrium.
This opens a new vista of unconventional phenomena that can be prepared and probed in OES.

As we were finalizing this manuscript, we became aware of the related work of Hu, Zoller, and Budich \cite{Hu2016}. Complementary to our work, they consider non-equilibrium Hall responses in the linear response regime in a ramp from a trivial to a topological phase.

\emph{Acknowledgements---}We thank Mehrtash Babadi, Eugene Demler, and Ian Spielman for helpful discussions. 
We thank the Air Force Office for Scientific Research (JW) and the Burke fellowship at Caltech (JCWS) for support. 
GR is grateful for support through the Institute of Quantum Information and Matter (IQIM), an NSF frontier center, supported by the Gordon and Betty Moore Foundation as well as the Packard Foundation and for the hospitality of the Aspen Center for Physics, where part of the work was performed.

\bibliography{references-arxiv.bib}

\onecolumngrid

\hspace{12pt}
\hrule
\hspace{12pt}

\section*{Supplementary Material to: \emph{Persistent Hall response in a quantum quench}}

\twocolumngrid

\section{General theory of quench-pulse protocol -- single particle}

In the single-particle framework, we work with a Hamiltonian diagonalized into (Bloch) wave vectors $h(\mathbf{k})$, so that $H = \sum_{\mathbf k} c_{\mathbf{k}}^\dagger h(\mathbf{k}) c_{\mathbf{k}}$.
We begin by preparing our state as the ground state of some Hamiltonian $h(\mathbf k, \lambda)$ where $\lambda$ is some parameter to be quenched by bringing it to zero at $t=0$.
At $t=0$ we quench to $h_0(\mathbf k) = h(\mathbf k,0)$, and we evolve for a time $t_1$ with this Hamiltonian.
At $t=t_1$ we then pulse the system with an electric field for a time shorter than time-scales in the problem, so that we can represent it by a shift in the Hamiltonian $h_0(\mathbf k - e \mathbf A) \equiv h_A(\mathbf k)$.
After evolution under this pulsed Hamiltonian for a time $t=t_2$, we then measure the current.

This whole process can be represented by the following final state
\begin{align}
  \ket{\psi_2} = e^{-i h_A(\mathbf k)(t_2 - t_1)} e^{-i h_0(\mathbf k) t_1} \ket{\psi_0},
\end{align}
where $h(\mathbf k, \lambda) \ket{\psi_0} = \epsilon \ket{\psi_0}$ (suppressing dependence on $\mathbf k$ and $\lambda$).  To easily identify the state right before we pulse, we define
\begin{align}
  \ket{\psi_1} = e^{-i h_0 t_1} \ket{\psi_0}.
\end{align} 

The current operator in single-particle quantum mechanics can easily be written as $\mathbf j_A = -e \bm \nabla_{\mathbf k} h_A$ (for the current when measured at time $t_2$).

In order to display the role of quantum geometry, we write the expectation value in a suggestive fashion
\begin{multline}
  \braket{\psi_2 | \mathbf j_A | \psi_2} = -e \nabla_{\mathbf k} \braket{ \psi_1 | h_A | \psi_1}  \\ 
  +e[\braket{\nabla_{\mathbf k} \psi_2 | h_A | \psi_2} +\braket{\psi_2 | h_A | \nabla_{\mathbf k} \psi_2} ]. \label{eq:current-hamiltonian}
\end{multline}
Here we introduce two objects, $E(\mathbf k;\lambda,t_1) = \braket{\psi_1|h_A|\psi_1}$ is the energy of the state after the pulse, and
\begin{align}
  F_{t_2,k_\mu} = i \braket{\partial_{t_2} \psi_2 | \partial_{k_\mu} \psi_2} - i \braket{\partial_{k_\mu} \psi_2 | \partial_{t_2} \psi_2}
\end{align}
is the Berry curvature of the evolving state (a function of $\mathbf k$, $\lambda$, $t_1$, and $t_2$). 
This can be introduced into the expression for current Eq.~\eqref{eq:current-hamiltonian} with use of the Schr\"odinger equation $i\partial_{t_2} \ket{\psi_2} = h_A \ket{\psi_2}$.

With these ingredients, we have (in coordinate representation)
\begin{align}
  \braket{\psi_2 | j_{A,\mu} | \psi_2} = -e(\partial_{k_\mu} E + F_{t_2 k_{\mu}}). \label{eq:current-energy-curvature}
\end{align}

We can now further investigate how this expression looks in terms of energies pre-quench.
In that case, we assume a linear dependence on $\mathbf k$ in our Hamiltonian (or if not, that $e|\mathbf A| \ll k$),
then we have
\begin{align}
  E & = \braket{\psi_1| h_0 - e A_\nu \partial_{k_\nu} h_0 | \psi_1}, \\
    & = E_0 - e A_\nu(\partial_{\nu} E_0 + F_{t_1 k_\nu}),
\end{align}
where we have defined $E_0 = \braket{\psi_0| h_0 | \psi_0}$ and $F_{t_1 k_\nu} = i \braket{\partial_{t_1} \psi_1 | \partial_{k_\nu} \psi_1} - i \braket{\partial_{k_\nu} \psi_1 | \partial_{t_1} \psi_1}$.

This allows us to write a general expression for the current as
\begin{multline}
  \braket{\psi_2 | j_{A,\mu} | \psi_2} = -e(\partial_{k_\mu} E_0 - e A_{\nu} \partial_{k_\mu} \partial_{k_\nu} E_0 \\
  -e A_\nu \partial_{k_\mu} F_{t_1 k_\nu} + F_{t_2 k_{\mu}}). \label{eq:current-curvatures}
\end{multline}

Note: This is approximate when the Hamiltonian is not linearly dependent on $\mathbf k$. Also, the Berry curvature $F_{t_2 k_\mu}$ depends strongly on $\mathbf A$. This somewhat complicated structure will manifest itself later when we consider the particle two-band model.

\subsection{Relation to the adiabatic anomalous velocity}

The derivation of Eq.~\eqref{eq:current-energy-curvature} did not depend on much except the use of the Schr\"odinger equation and energy conservation. 
However, for time-dependent Hamiltonians we have 
\begin{align}
  \braket{\psi(t) | j_\mu | \psi(t)} = -e[ \partial_{k_\mu} E(t) + F_{t k_\mu} ].
\end{align}
To see how this expression relates to anomalous velocities in band structure, consider that $\ket{\psi(t)}$ satisfies the adiabatic theorem for $h_0( \mathbf k - e \mathbf A(t)) \ket{\psi(t)} = \epsilon(\mathbf k- e \mathbf A(t)) \ket{\psi(t)}$.
As an immediate consequence, we can expand $E(t) = \epsilon(\mathbf k - e \mathbf A(t)) \approx \epsilon(\mathbf k) - e  A_\nu \partial_{k_\nu} \epsilon(\mathbf k)$.

Additionally, in linear response and ignoring an overall phase factor that can be gauged away, $\ket{\psi(t)} = \ket{\epsilon(\mathbf k)} - e A_\nu (t) \partial_{k_\nu} \ket{\epsilon(\mathbf k)} + \cdots $. Thus, $\partial_t \ket{\psi(t)} = - e \dot A_\nu \ket{ \partial_{k_\nu} \epsilon}$, and we can write the Berry curvature as
\begin{align}
  F_{t k_\mu} = - e \dot A_{\nu}(t) \Omega_{k_\nu k_\mu},
\end{align}
where the Berry curvature $\Omega_{k_\nu k_\mu}$ is entirely in the ground-state manifold.

Combining these, we have
\begin{align}
  \braket{j_\mu} = -e[\partial_{k_\mu} \epsilon - e \partial_{k_\mu} A_{\nu} \partial_{k_\nu} \epsilon - e \dot A_{\nu}(t) \Omega_{k_\nu k_\mu}].
\end{align}

We can rewrite this, recalling that $\mathbf j = - e \dot {\mathbf x}$, and making a vector out of $\Omega$ with $\bm \Omega = \frac12 \epsilon^{\alpha \beta \gamma} \Omega_{k_\beta k_\gamma} \hat{\mathbf{e}}_\alpha $.
The end result is

\begin{align}
  \braket{\dot{\mathbf x}} = \nabla_{\mathbf k} \epsilon - e\nabla_{\mathbf k} (\mathbf A(t) \cdot \nabla_{\mathbf k}\epsilon) - e \bm \Omega \times \dot{\mathbf A}(t).
\end{align}

The first term $\nabla_{\mathbf k} \epsilon$ is the group velocity, the second term $- e\nabla_{\mathbf k} (\mathbf A(t) \cdot \nabla_{\mathbf k}\epsilon)$ is the ballistic response, and $- \tfrac e 2 \bm \Omega \times \dot{\mathbf A}(t)$ is the anomalous velocity which came directly from $F_{t k_\mu}$. We note that the ballistic response is sensitive to scattering such as scattering of impurities.

\subsection{Persistent currents}

When a system is pulsed, it is possible to develop a persistent current that, without dissipation, lasts long after the pulse. In the simplest situation, the free particle retains velocity after a momentum kick.

Returning to the early expression for current $\braket{\psi_2 | \mathbf j_A |\psi_2}$, we can expand $\ket{\psi_2}$ in terms of eigenstates of the new pulsed Hamiltonian $h_A(\mathbf k)$, in which case
\begin{align}
  \ket{\psi_2} = \sum_n \ket{\epsilon_{A,n}} \braket{\epsilon_{A,n} | \psi_1} e^{-i \epsilon_{A,n} (t_2 - t_1)}
\end{align}
where $h_A \ket{\epsilon_{A,n}} = \epsilon_{A,n} \ket{\epsilon_{A,n}} $.  This implies that 
\begin{multline} \label{eq:current-eigenstates}
  \braket{\psi_2 | \mathbf j_A |\psi_2 } = \sum_{n,m} \braket{\psi_1 | \epsilon_{A,n}} \braket{\epsilon_{A,n} | \mathbf j_A | \epsilon_{A,m}} \\ \times \braket{\epsilon_{A,m}|\psi_1} e^{-i (\epsilon_{A,m} - \epsilon_{A,n})(t_2 - t_1)}.
\end{multline}
Now, to pick out what will be persistent, we take what this current is oscillating around, so the only component we keep is $n = m$. This crucially depends on the system having finite bands with finite energy gaps.
\begin{align}
  \braket{\mathbf j_A(t_2 \rightarrow \infty )} = \sum_{n}  \braket{\epsilon_{A,n} | \mathbf j_A | \epsilon_{A,n}} |\braket{\psi_1 | \epsilon_{A,n}}|^2 .
\end{align}
And for eigenstates, we have that 
\begin{align}
  \braket{\epsilon_{A,n} | \mathbf j_A | \epsilon_{A,n}} & = \braket{\epsilon_{A,n} | \nabla_{\mathbf k} h_A | \epsilon_{A,n}} \\
   & = \nabla_{\mathbf k} \braket{\epsilon_{A,n} |  h_A | \epsilon_{A,n}} = \nabla_{\mathbf k} \epsilon_{A,n}.
\end{align}
Thus, we have
\begin{align}
  \braket{\mathbf j_A(t_2 \rightarrow \infty )} = \sum_{n}  (\nabla_{\mathbf k} \epsilon_{A,n}) |\braket{\psi_1 | \epsilon_{A,n}}|^2 .
\end{align}

This obscures any play with quantum geometry which the previous sections elucidated. 

If we take this one step further to when we perform the quench, we can write the expression as 
\begin{align}
  \braket{\mathbf j_A(t_2 \rightarrow \infty )} = \sum_{n}  (\nabla_{\mathbf k} \epsilon_{A,n}) |\braket{\psi_0 | e^{i t_1 h_0 }| \epsilon_{A,n}}|^2 . \label{eq:persistent-current-gen-expr}
\end{align}

In order to measure the Hall current, we take $\mathbf A$ and $\mathbf k$ perpendicular to each other.

\subsection{Right after the pulse}

If we measure current directly after the pulse, the state has not had time to evolve and the current will be
\begin{align}
  \braket{\psi_1 | j_{A,\mu} | \psi_1} & \approx \braket{\psi_1 | \mathbf j_{0,\mu} | \psi_1} 
  + A_\nu \braket{\psi_1 | \partial_{k_\mu}\partial_{k_\nu} h_0 | \psi_1}. \label{eq:short-time-current-singlestate}
\end{align}
For a Hamiltonian linear in $\mathbf k$, there is no immediate response for the single particle state. 

\section{The ``Haldane'' model}

We now consider the specific case of the Haldane model. 
In this model, there are two Dirac cones in a 2D Brillouin zone where one of the cones is (near the Dirac point)
\begin{align}
  h(\mathbf k,\Delta) = v_{\mathrm F} \mathbf k \cdot \sigma + \Delta \sigma_z.
\end{align}
The second cone is similar except the kinetic energy term is related by the time-reversal operator $T=K$ (complex conjugation), but time-reversal symmetry is broken so $\Delta\rightarrow-\Delta$ as well.
Breaking of time-reversal symmetry is essential, otherwise the Hall response we witness below is just zero. 

This procedure is detailed in the following section. We (1) prepare the state in $h(\mathbf k, \Delta)$, (2) then at $t=0$ we quench to $h(\mathbf k,0)$, (3) pulse at $t=t_1$ to $h(\mathbf k - e\mathbf A,0)$, and (4) measure current at $t=t_2$.

\subsection{State preparation}

Beginning with $h(\mathbf k, \Delta)$, we can prepare the state in the ground state.  
This state is represented is represented by
\begin{align}
  \ket{\psi_0} = \cos\tfrac\theta2 \ket{\uparrow} - e^{i\phi} \sin\tfrac\theta 2 \ket{\downarrow},
\end{align}
where
\begin{align}
  \cos\theta &= -\frac{\Delta}{\sqrt{k^2 + \Delta^2}}, \\
  k e^{i\theta} & = k_x + i k_y.
\end{align}

\subsection{Quench and evolve}

With this state, we quench into $h_0(\mathbf k, 0)$. 
The time evolution operator, as can be checked, is
\begin{align}
  e^{-i t_1 h_0} & = \cos k t_1 - i \sin k t_1 \frac{\mathbf k \cdot \sigma}k \\
    & = \begin{pmatrix}
      \cos k t_1 &  -i \sin k t_1 e^{-i\phi} \\
      -i \sin k t_1 e^{i \phi} & \cos k t_1
    \end{pmatrix}.
\end{align}
Thus, our state evolves to 
\begin{align}\label{eq:quench-evolved-state}
  \ket{\psi_1} = f(t_1)\ket{\uparrow} - e^{i\phi} g(t_1) \ket{\downarrow},
\end{align}
where
\begin{align}
  f(t_1) & = \cos\tfrac\theta2 \cos k t_1 + i \sin\tfrac\theta2 \sin k t_1, \\
  g(t_1) & = \sin\tfrac\theta 2\cos k t_1 + i \cos\tfrac\theta2 \sin k t_1.
\end{align}

\subsection{Pulse}

We now consider pulsing the state.
The new evolution operator is the same as the above let $\mathbf k \rightarrow \mathbf k - e \mathbf A$.

In order to obtain analytically tractable solutions, we assume that $e |\mathbf A| \ll \mathbf k$. 
Without loss of generality, we take the pulse in the $x$-direction, and the above expansion implies 
\begin{align}
  | \mathbf k - e \mathbf A| & = k - e A_x \cos\phi, \\
  e^{i (\phi_A - \phi)} & = 1 + i \tfrac{e A_x}{k} \sin\phi,
\end{align}
where $e^{i \phi_A} = (k_x - eA_x) + i k_y$.
The time evolution operator can then be appropriately expanded while making \emph{no assumptions} regarding the time:
\begin{widetext}
\begin{align}\label{eq:pulse-evolve}
  e^{- i \delta t \, h_A} & = \begin{pmatrix}
     \cos[(k - e A_x \cos\phi)\delta t] & - i\sin[(k - e A_x \cos\phi)\delta t](1 - i \tfrac{eA_x}{k} \sin\phi)  e^{-i \phi} \\
     -i \sin[(k - e A_x\cos\phi) \delta t](1 + i \tfrac{e A_x}{k}\sin\phi) e^{i\phi} & \cos[(k - e A_x \cos\phi) \delta t]
  \end{pmatrix}.
\end{align}
\end{widetext}
Here, time $\delta t = t_2 - t_1$.
Now, we can evolve our state with $\ket{\psi_2} = e^{-i(t_2 - t_1) h_A} \ket{\psi_1}$. 
The resulting expression can be deduced by matrix multiplication of Eq.~\eqref{eq:pulse-evolve} and Eq.~\eqref{eq:quench-evolved-state}.

\subsection{Measuring the current}

We are interested in the current for the entire system where the $\mathbf k$ states in the valence band of $h(\mathbf k, \Delta)$ are fully occupied. This analysis can also be applied for filled states up to some chemical potential $\mu$.
In this section, we build up to that by looking first at the current of an individual state at $\mathbf k$ in the valence band, then of a ring of states with the same $|\mathbf k|$, and finally at the entire band itself.

\subsubsection{The current of a single state}

Having constructed the single-particle state $\ket{\psi_2}$, we can calculate the expectation value of $j_y = - e \sigma_y$ with the use of
\begin{widetext}
\begin{multline}
  \braket{\psi_2 | \sigma_y |\psi_2} = \Im e^{i\phi} \left\{-i(\lvert f(t_1)\rvert^2 - \lvert g(t_1)\rvert^2) \left(1+ i\tfrac{e A_x}{k}\sin\phi \right) \sin{\left [2 (k - e A_x \cos\phi)(t_2-t_1) \right ]}  \right. \\ 
  - f(t_1)^* g(t_1)(1 + \cos{\left [2 (k - e A_x \cos\phi)(t_2-t_1) \right ]})  \\ \left.
  - g^*(t_1) f(t_1) \left(1+2 i \tfrac{e A_x}{k}\sin\phi\right) (1 - \cos{\left [2 (k - e A_x \cos\phi)(t_2-t_1) \right ]}) \right \},
\end{multline}
\end{widetext}
and it is easily shown that
\begin{align}
  | f(t_1)|^2 - | g(t_1)|^2 & = \cos\theta \cos  2 k t_1 ,\\
   f(t_1)^* g(t_1) & = \tfrac12(\sin\theta +i \cos\theta \sin 2 k t_1).
\end{align}

Notice how the persistent current previously alluded to is already showing up.
If one drops out all oscillating currents and terms that do not depend on $A_x$, we obtain a term $2 \tfrac{e A_x}p \sin\phi \Re e^{i\phi} g^* f$.
This term will continue throughout the calculations, giving the persistent effect.

\begin{figure}[t!]
\centering
\subfloat[Single state current response for $p=3\Delta/v_{\mathrm F}$, $\phi=0$, and $e A_x = 0.3\Delta/v_{\mathrm F}$.]{
\includegraphics[width=\columnwidth]{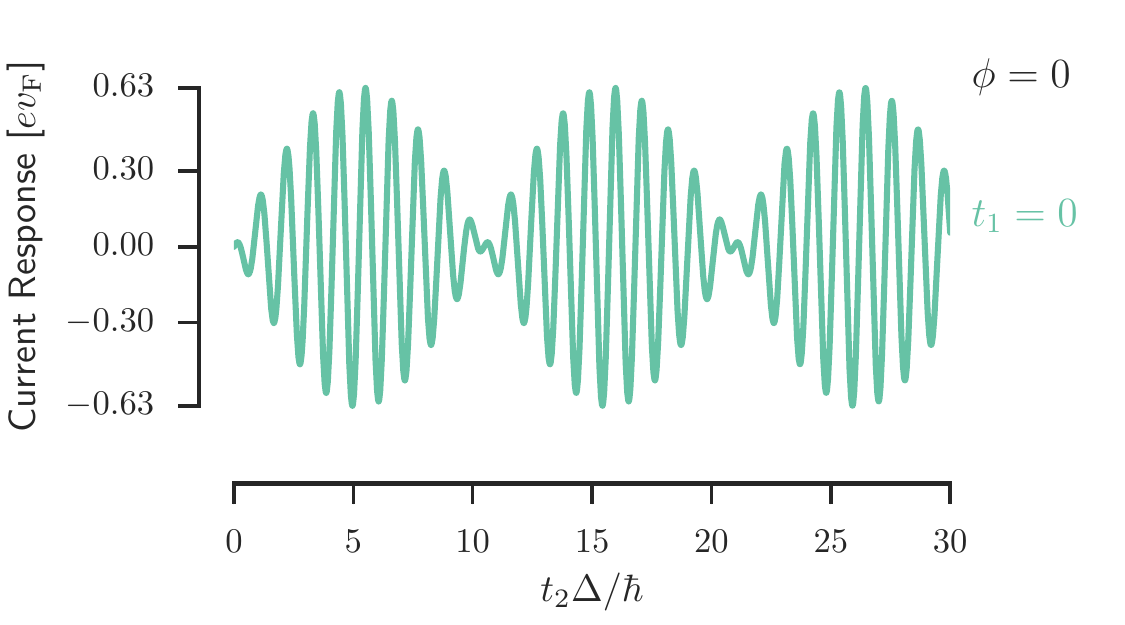}
\label{fig:single-particle-current-phizero}
}

\subfloat[Single state current response for $p=3\Delta/v_{\mathrm F}$, $\phi=\pi/2$, and $e A_x = 0.3\Delta/v_{\mathrm F}$.]{
\includegraphics[width=\columnwidth]{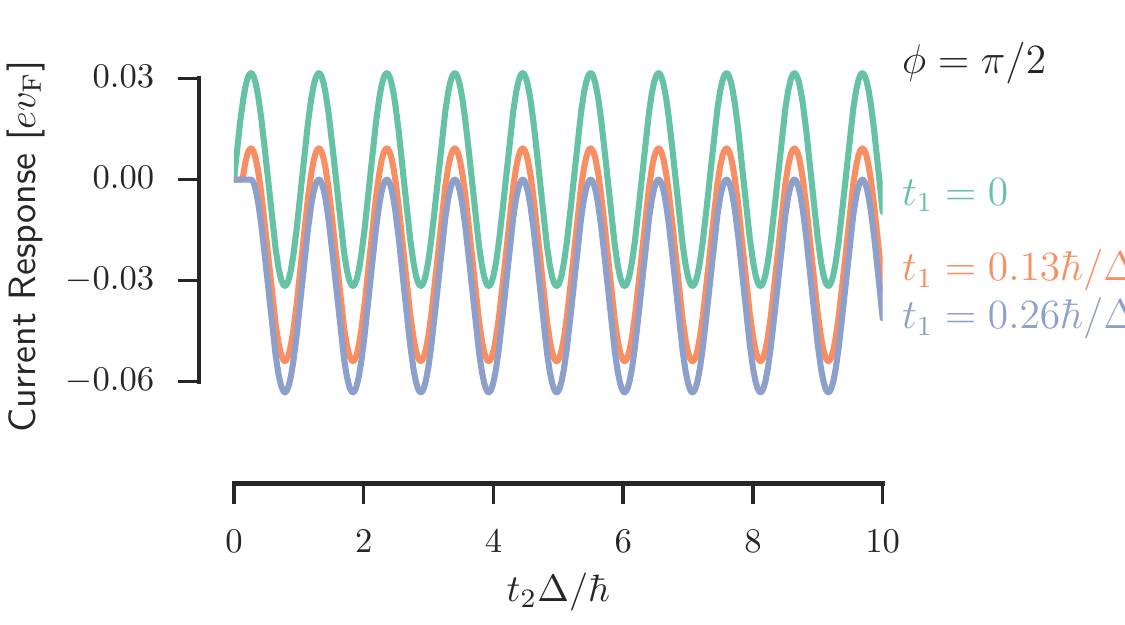}
\label{fig:single-particle-current-phihalfpi}}
\caption{The single-particle current response: $\braket{j_y} - \braket{j_y}|_{A_x=0}$ using state $\ket{\psi_2}$. For \protect\subref{fig:single-particle-current-phizero}, the state's momentum is parallel to the pulse and we see the resulting beating due to a change in frequency. On the other hand, in \protect\subref{fig:single-particle-current-phihalfpi} we see no sign of beating but the center of oscillation is shifted from zero for different pulse times $t_1$.}
\label{fig:single-particle-current}
\end{figure}

We note the current is oscillatory. 
However, to understand how it is deviating from the non-pulsed current, we plot in Fig.~\ref{fig:single-particle-current} the difference $\braket{\sigma_y} - \braket{\sigma_y}|_{A_x\rightarrow 0}$.
In this figure, note that the pulse can change the frequency, leading to a beating effect. But for states that have momentum perpendicular to the pulse ($\phi = \pi/2$), the center of oscillation is shifted.

\subsubsection{The current of states with the same momentum}

Now, we take the above and integrate it around a ring of constant of momentum. 
In particular, $j_y(k) = -e \int_0^{2\pi} \frac{d\phi}{2\pi} \braket{\psi_2(\mathbf k)| \sigma_y | \psi_2(\mathbf k)}$, and we obtain
\begin{widetext}
  \begin{align}
  j_y(k) = -\frac{\Delta}{\sqrt{k^2 + \Delta^2}} \left\{ \frac{e^2 A_x}{2k} \sin 2 k t_1  - e J_1[2 e A_x(t_2-t_1)] \left( \cos 2k t_2 + \frac{\sin 2 k t_2}{2k(t_2 - t_1)}\right) \right\}. \label{eq:ring-current}
\end{align}
\end{widetext}

\begin{figure}[t!]
\includegraphics[width=\columnwidth]{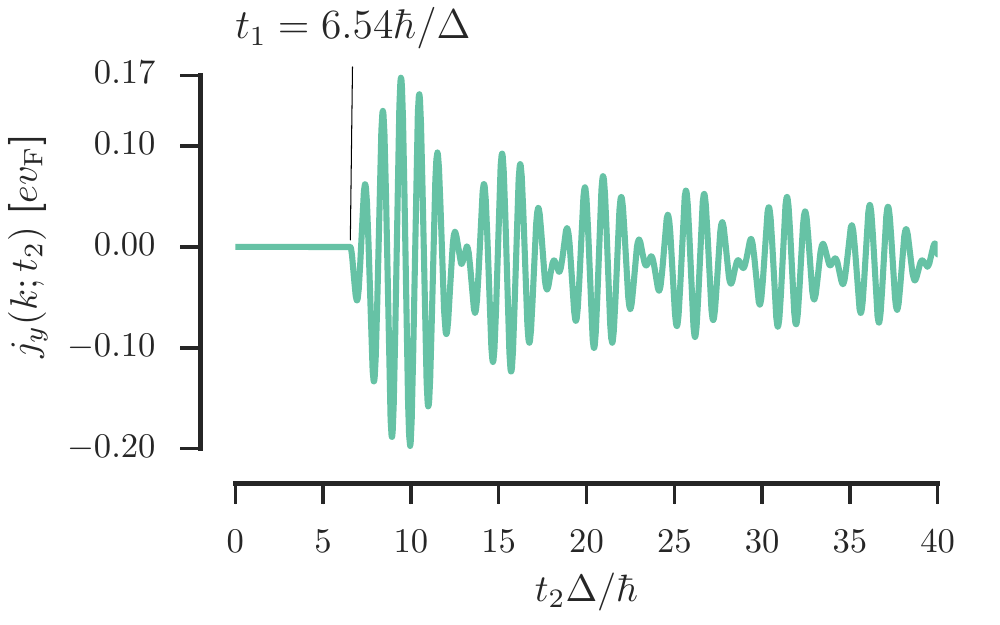}

\caption{The combined current for all states at a constant momentum $p = 3\Delta/v_{\mathrm F}$ and $e A_x = 0.3\Delta/v_{\mathrm F}$. We illustrate the Bessel function behavior at a finite pulse-time $t_1$. While difficult to discern, this still oscillates about a finite, persistent value.}
\label{fig:ring-current}
\end{figure}

We still have a persistent effect, and in addition we have a term that dies off with $t_2$ (the Bessel function).
This is plotted in Fig.~\ref{fig:ring-current}.

\subsubsection{The current of the whole band}

We now find the total current $J_y(t_2;t_1)$ for states originally loaded entirely in the valence band of the Haldane model.
We have previously assumed $k\ll e A_x$, but here we will assume $\Delta/v_{\mathrm F} \ll e A_x$ since we will be integrating over all $k$. 

To find the total current then, we need to evaluate 
\begin{align}
  J_y & = \int_0^\infty \frac{k \, dk}{2\pi} j_y(k).
\end{align}

Inspection of Eq.~\eqref{eq:ring-current} reveals that there are only two integrals to consider and we consider them in turn:
\begin{align}
 A_1 & = \int_0^\infty \frac{k \, dk}{2\pi|\Delta|} \frac{\cos 2 k t}{\sqrt{k^2 + \Delta^2}}, \\
 A_2 & = \int_0^\infty \frac{dk}{2\pi} \frac{\sin 2 k t}{\sqrt{k^2 + \Delta^2}}.
\end{align}
It seems as though $A_1$ is infinity, but we can evaluate it as
\begin{align}
 A_1 & = \int_0^\infty \frac{x dx}{2\pi} \frac{\cos 2 x t |\Delta|}{\sqrt{x^2 + 1}} \\
   & = \int_0^\infty \frac{dx}{2\pi} \cos 2 x t |\Delta| + \int_0^\infty \frac{dx}{2\pi} 
   \frac{\cos 2 x t |\Delta|}{x^2 + 1 - x\sqrt{x^2 + 1}}.
\end{align}
The first term is purely oscillatory and can be evaluated as a $\delta$-function.
The second term can be rewritten as a convergent integral using complex analysis;
\begin{align}
  A_1 = \tfrac12 \delta(2 t |\Delta|) - \int_0^{\pi/2} \frac{dz}{2\pi} \sin z \,e^{-2 t |\Delta| \sin z}.
\end{align}
On the other hand, $A_2$ can be easily converted as well into
\begin{align}
  A_2 = \int_0^{\pi/2} \frac{dz}{2\pi} e^{-2 t |\Delta| \sin z}.
\end{align}
In terms of modified Bessel functions ($I_\nu(x)$) and modified Struve functions ($L_\nu(x)$), we can write these as
\begin{align}
  A_1 & = \tfrac12\delta(2t|\Delta|) +\tfrac14(I_1(2t|\Delta|) - L_{-1}(2t|\Delta|)),\\
  A_2 & = \tfrac14(I_0(2t|\Delta|) - L_{0}(2t|\Delta|)).
\end{align}
The $\delta$-function will not be an issue since it has support only when $t_2 = 0$. 

For ease of notation, we define
\begin{align}
  K_\nu(x) = \int_0^{\pi/2} \frac{dz}{2\pi} (-1)^\nu \sin^\nu z \, e^{-2 x \sin z}.
\end{align}

Thus, the overall current can be written as (putting in all the appropriate units)
\begin{widetext}
\begin{align}
  J_y =  -\frac{\pi e^2}{h}\left\{ \frac{A_x \Delta}{\hbar}
     K_0(t_1 \tfrac{|\Delta|}{\hbar}) - J_1[2 v_{\mathrm F} eA_x(t_2-t_1)/\hbar] \left[\frac{\Delta}{e v_{\mathrm F}(t_2 - t_1)} K_0(t_2 \tfrac{|\Delta|}{\hbar}) + \sgn(\Delta) \frac{\Delta^2}{e v_{\mathrm F} \hbar}K_1(t_2 \tfrac{|\Delta|}{\hbar}) \right]\right\}.
\end{align}
\end{widetext}
This form of $J_y$ allows us to investigate some of the asymptotic properties.
Importantly, the first term is clearly what we get as a persistent current for 
$t_2 \rightarrow \infty$.
Additionally, in that same limit we can use the asymptotic form of $J_1$ yielding the next order term that dies off as $t_2^{-5/2}$.

The current is zero when $t_1 = t_2$. However, at long times, we can see the persistent effect $J_y^\infty$ which we explicitly define
\begin{align}
  J_y^\infty = -\frac{\pi e^2}{h}\left(\frac{A_x \Delta}{\hbar}\right)
     K_0(t_1 \tfrac{|\Delta|}{\hbar}). \label{eq:Jy-infinity-approx}
\end{align}
As $t_1\rightarrow 0$, we get
\begin{align}
  J_y^\infty \sim -\frac{\pi e^2}{4h}\left(\frac{A_x \Delta}{\hbar}\right) \left( 1 - \frac{t_1 |\Delta|}{\pi\hbar}\right).
\end{align}
On the other hand, as $t_1\rightarrow \infty$
\begin{align}
  J_y^\infty \sim - \sgn(\Delta) \frac{e^2}{4h} \frac{A_x}{t_1}.
\end{align}
This die off with $t_1$ is due to dephasing in the system.

\begin{figure}[b!]
\includegraphics[width=\columnwidth]{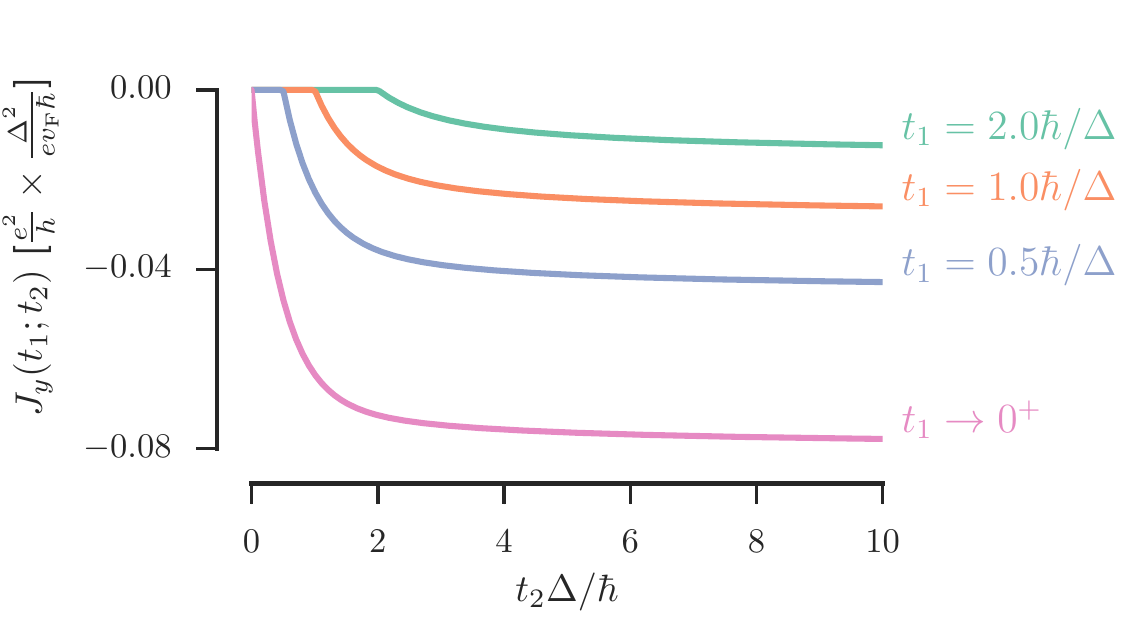}
\caption{The current of the whole valence band of the Haldane model following the quench and subsequent pulse with size $e A_x = 0.1 \Delta/v_{\mathrm F}$. The current starts at zero until it is pulsed at $t_1$ at which point it moves towards its persistent value $J_y^\infty$.}
\label{fig:total-current}
\end{figure}

The current itself can be seen for various values of $t_1$ in Fig.~\ref{fig:total-current}.
Note that the current starts at zero and grows to saturate at its persistent value.
We note that the order of limits matters here: If we let $t_1 = 0$ before we integrate, we would not get any persistent value;
however, if we wait any amount of time, we immediately get the large persistent value here.
This can be understood by how the states are evolving on the Bloch sphere as we will discuss in a later section and in the main text.

With all of this, we now look to analyze explicitly the persistent current. In so doing, we will use to quantum geometry as a tool. 

\section{Bloch geometry and the persistent Hall current}

Now we consider the Bloch sphere and what it means to have a persistent current.
In our example, the current is given exclusively by $\braket{\sigma_y}$ which is just the projection of our Bloch vector onto the $y$-axis.

\begin{figure*}
  \includegraphics{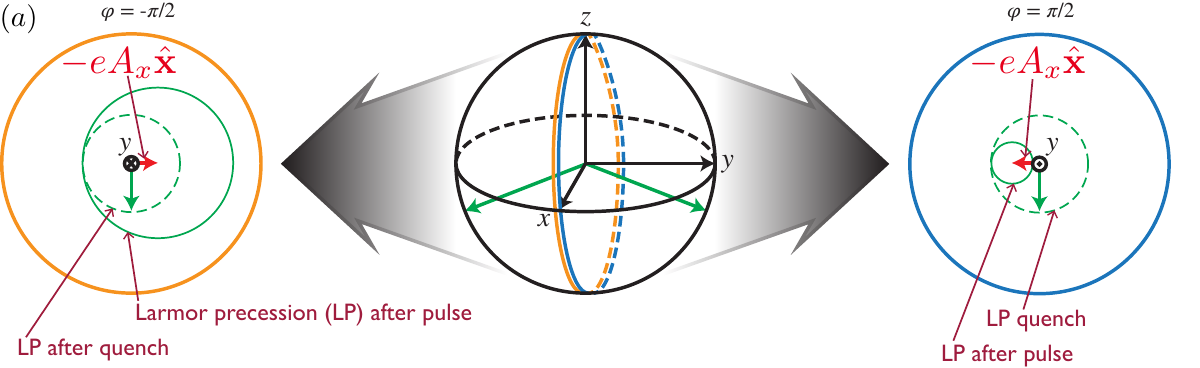}
  \caption{The green arrows represent the states being considered: Both have constant $p$ and are at angles $\phi=\pm \pi/2$. The pulse causes the states to rotate clockwise around a point north of it on the equator (it rotates around the momentum vector $\mathbf p$). At time $t_1 = \frac{\pi}{8 k v_\mathrm{F}}$, both states are on the equator and we apply a pulse $e A_x> 0$. This shifts the center of rotation making the circle representing Rabi oscillation smaller for $\phi = \pi/2$ and larger for $\phi = -\pi/2$. In Fig.~\ref{fig:BlochSphereExplanation2} we show how this leads to a shift in the average $\braket{\sigma_y}$.}
  \label{fig:BlochSphereExplanation1}
\end{figure*}

\begin{figure}
  \includegraphics{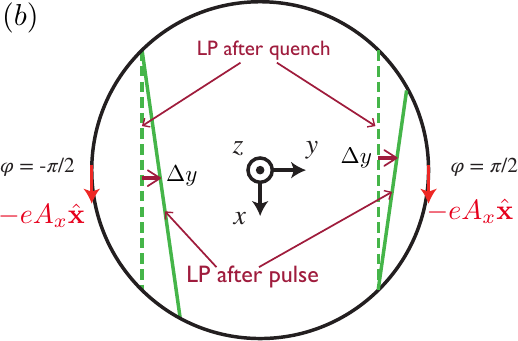}
  \caption{From a bird's eye-view, we can see that the Rabi oscillations represented in Fig.~\ref{fig:BlochSphereExplanation1} cause a movement in the center of rotation along the $y$-direction (represented by $\Delta y$ in the figure). Just as in Fig.~\ref{fig:BlochSphereExplanation1}, we have taken $p$ the same for both and $t_1 = \frac{\pi}{8 k v_\mathrm{F}}$.}
  \label{fig:BlochSphereExplanation2}
\end{figure}

The idea is captured in Figures \ref{fig:BlochSphereExplanation1} and \ref{fig:BlochSphereExplanation2}.
The key feature being: the pulse shifts the center of rotation, in part, perpendicular to the pulse.
Therefore, we get an average current in that direction that is seen once the system dephases.

This persistent current can be calculated exactly using these ideas \emph{without} the approximation $e A_x \ll k$ that we made in previous sections.


\subsection{Bloch geometry}

Say we have a path on the Bloch sphere $\mathbf r(t)$ which rotates around some vector $\mathbf B$, then the average of this path around one cycle can be easily given by a projection $(\mathbf r(t) \cdot \hat{\mathbf B}) \hat{\mathbf B}$ where $\hat{\mathbf B} = \mathbf B/B$ is a unit vector.
If we further want to know how much of this is along the $y$-direction, we can project it along $y$, giving us simply
\begin{align}
  Y(\hat{\mathbf B}; \mathbf r(t)) \equiv (\mathbf r(t) \cdot \hat{\mathbf B})(\hat{\mathbf{B}} \cdot \hat{\mathbf y}). \label{eq:blochsphereprojection}
\end{align}
This quantity represents the persistent current for a particular state and is independent of $t$ (given $\mathbf r(t)$ circulating around $\hat{\mathbf{B}}$), and when we integrate it over all states we will obtain the persistent current as previously discussed.

After our mapping to the Bloch sphere, we begin in a state on there sphere $\mathbf r_0$, rotate into $\mathbf r_1$ at time $t_1$ by precession around the vector $2\mathbf p$. The pulse then causes precession about the vector $\mathbf B = 2 (\mathbf p - e \mathbf A)$, but averaging the result over a period of precession gives exactly Eq.~\eqref{eq:blochsphereprojection} with $\mathbf r(t) = \mathbf r_1$.
Therefore, to find $\mathbf r_1$, we look at the precession of our initial state $\mathbf r_0$ around the vector $2\mathbf p$ ($\ket{\psi_i}$ is represented by Bloch vector $\mathbf r_i$ here). Mapping our vector appropriately,
\begin{align}
  \mathbf r_0 = - \hat{\mathbf p} \sin\theta + \hat{\mathbf z} \cos\theta.
\end{align} 
As we rotate around $\mathbf{p}$ at a rate of $2p$, we just need to let $\hat{\mathbf z} \rightarrow (\cos 2 p t \, \hat{\mathbf z} + \sin 2 p t \, \hat{\mathbf{z}} \times \hat{\mathbf p})$. Therefore,
\begin{align}
  \mathbf r_1 = - \hat{\mathbf p} \sin\theta + (\cos 2 p t_1 \, \hat{\mathbf z} - \sin 2 p t_1 \, \hat{\mathbf{z}} \times \hat{\mathbf p})\cos\theta.
\end{align}
The quantity we are interested in for this problem is
\begin{align}
  Y(p,\phi) = \frac{[\mathbf r_1 \cdot (\mathbf p - e\mathbf A)][(\mathbf p - e \mathbf A)\cdot \hat{\mathbf y}]}{|\mathbf p - e \mathbf A|^2}.
\end{align}
Imposing $\mathbf A = A_x \hat{\mathbf x}$, we can begin evaluating
\begin{align}
  (\mathbf p - e \mathbf A)\cdot \hat{\mathbf y} = p \sin \phi.
\end{align}
Then, in order to evaluate $\mathbf r_1 \cdot (\mathbf p - e \mathbf A)$ we find
\begin{align}
  \hat{\mathbf p} \cdot (\mathbf p - e \mathbf A) & = p - e A_x\cos\phi, \\
  (\hat{\mathbf z} \times \hat{\mathbf p}) \cdot (\mathbf p - e\mathbf A) & =  e A_x \sin \phi.
\end{align}
These equations imply
\begin{align}
  Y(p,\phi) = \tfrac{p[-\sin\theta(p - e A_x \cos\phi)- e A_x \sin 2 p t_1 \, \cos\theta \, \sin\phi]\sin\phi}{p^2 + e^2 A_x^2 - 2 e A_x p \cos \phi}.
\end{align}
Just as before, we integrate first around $d \phi$.
With some complex integration, we can write the final result as
\begin{align}
  \int\frac{d\phi}{2\pi} Y(p,\phi) = -\frac12 \cos\theta \sin 2 p t_1 \begin{cases}
  \frac{e A_x}{p}  & |e A_x| < p, \\
  \frac{p}{ e A_x} & |e A_x| > p.
  \end{cases}
\end{align}
And finally, we can integrate this expression over momentum to obtain $J_y^\infty$. Inserting all physical constants:
\begin{widetext}
\begin{align}
  J_y^\infty = -\frac{e^2}{2h} \left(\frac{A_x \Delta}{\hbar}\right)\left[ \int_0^{\pi/2} dz \,e^{-2 t_1 |\Delta|\sin z} +
    \frac{|e A_x|}{\Delta} \int_0^1 dx \frac{x^2 - 1}{\sqrt{1+(\frac{eA_x x}{\Delta})^2}} \sin(2 |e A_x| x t_1) \right]. 
    \label{eq:Jy-infinity-exact}
\end{align}
\end{widetext}
At lowest order Eq.~\eqref{eq:Jy-infinity-exact} agrees with Eq.~\eqref{eq:Jy-infinity-approx}, and we see that the higher order terms do in fact die off as $eA_x/\Delta$.

\subsection{General two-band theory}

We now consider the general two-band theory for this Hall effect.
In general, we define the persistent Hall response as the antisymmetric part of the tensor $\Sigma_{\mu \nu}^\infty$ that is obtained from
\begin{align}
  J_\mu^\infty = \Sigma_{\mu \nu}^\infty A_\nu + O(A^2).
\end{align}
Since we are working in two-dimensions, this is just a single value we call $\Sigma_{\rm Hall}^\infty = \Sigma_{yx}^\infty - \Sigma_{xy}^\infty$.

The general two band model we have in mind is
\begin{align}
  h(\mathbf p, \Delta) = \epsilon_0(\mathbf p) \mathbb I + \mathbf d(\mathbf p, \Delta)\cdot \bm \sigma.
\end{align}
Without loss of generality, we quench $\Delta \rightarrow 0$ at $t=0$. The current operator for such a theory is $\partial_\mu h(\mathbf p, \Delta) = \partial_\mu \epsilon_0(\mathbf p) \mathbb I + \partial_\mu \mathbf d(\mathbf p, \Delta)$.

First, let us show that $\partial_\mu \epsilon_0(\mathbf p) \mathbb I$ does not contribute to the Hall conductivity. Considering the state $\ket{\psi_2}$, the current after the pulse is
\begin{align}
  \braket{\psi_2 |\partial_\mu \epsilon_0(\mathbf p- e \mathbf A) | \psi_2} &= \partial_\mu \epsilon_0(\mathbf p- e \mathbf A) \\ &= \partial_\mu \epsilon_0(\mathbf p)- e A_\nu\partial_\mu \partial_\nu \epsilon_0(\mathbf p).
\end{align}
The last term is symmetric in $\mu$ and $\nu$ and therefore cannot contribute to a Hall response.
Therefore, we only need to consider $\partial_\mu \mathbf d(\mathbf p, \Delta) \cdot \bm \sigma$ for the \emph{Hall} current.

Taking the view from before, we have simply that the current at infinite time should be represented by
\begin{align}
  j^\infty_\mu(\mathbf p, t_1) = -e \partial_\mu \mathbf d(\mathbf p - e\mathbf A, \Delta) \cdot \overline{\braket{\psi_2 | \bm \sigma | \psi_2}},
\end{align}
where $\overline{\braket{\psi_2 | \bm \sigma | \psi_2}}$ represents the time-average of the state over a period of Larmor precession. 

The average is purely determined by the state right after the pulse $\braket{\psi_1 | \bm \sigma | \psi_1}$ which is a point on the Bloch sphere, and this is determined by the evolution of $\hat{\mathbf n} = \braket{\psi(t) | \bm \sigma | \psi(t)}$ in the time frame $t\in[0,t_1)$.
The equation of motion is
\begin{align}
  \partial_{t} \hat{\mathbf n} = 2 \mathbf d(\mathbf p, 0) \times \hat{\mathbf n}, \quad \hat{\mathbf n}(t=0) = -\hat{\mathbf d}(\mathbf p, \Delta). \label{eq:LarmorPrecession1}
\end{align}

After the pulse, the new center of rotation is defined by the vector $\mathbf d(\mathbf p - e \mathbf A)$, and the average is simply
\begin{align}
  \overline{\braket{\psi_2 | \bm \sigma | \psi_2}} = \hat{\mathbf d}(\mathbf p - e \mathbf A,0) [\hat{\mathbf d}(\mathbf p - e \mathbf A,0) \cdot \hat{\mathbf n}(t_1)].
\end{align}

It is then a simple matter to show
\begin{align}
  j^\infty_\mu(\mathbf p, t_1) =  -e[\hat{\mathbf d}(\mathbf p - e \mathbf A,0) \cdot \hat{\mathbf n}(t_1)] \partial_\mu d(\mathbf p- e \mathbf A, 0).
\end{align}

The integral of this over the occupied momenta $\mathbf p$ will give the total current at infinite-time $J_\mu^\infty$. 
However, we are interested in picking out the Hall contribution.
We find that if we define $j_\mu^\infty(\mathbf p, t_1) \approx \chi_{\mu \nu}^\infty(\mathbf p, t_1) A_\nu$ and $ \chi_{\rm Hall}^\infty(\mathbf p, t_1) = \frac12(\chi_{yx}^\infty - \chi_{xy}^\infty) $, then $\Sigma_{\rm Hall}^\infty = \int_p \chi_{\rm Hall}^\infty$, $\hat{\mathbf d}_0 = \hat{\mathbf d}(\mathbf p, 0) $, and $\hat{\mathbf d} = \hat{\mathbf d}(\mathbf p, \Delta) $, and
\begin{align}
  \chi_{\rm Hall}^\infty(\mathbf p, t_1) = e^2 [  \partial_y d_0 \partial_x \hat{\mathbf d}_0 - \partial_x d_0 \partial_y \hat{\mathbf d}_0 ] \cdot \hat{\mathbf n}(t_1). \label{eq:persistentHall-to-integrate}
\end{align}

\subsubsection{Condition for no persistent Hall current}

\newcommand{\dd}{\hat{\mathbf{d}}}
We can solve Eq.~\eqref{eq:LarmorPrecession1} and obtain
\begin{multline}
  \hat{\mathbf n}(t_1) = \dd_0(\dd \cdot \dd_0) - \dd_0 \times(\dd_0 \times \dd) \cos 2 d_0 t_1 \\ - \dd_0 \times \dd \sin 2d_0 t_1.
\end{multline}
This can be substituted into Eq.~\eqref{eq:persistentHall-to-integrate}, and we get
\begin{multline}
  \chi_{\rm Hall}^\infty(\mathbf p, t_1) = e^2 [  \partial_y d_0 \partial_x \hat{\mathbf d}_0 - \partial_x d_0 \partial_y \hat{\mathbf d}_0 ] \\ \cdot [\dd  \cos 2 d_0 t_1  - \dd_0 \times \dd \sin 2d_0 t_1]. \label{eq:persistentHall-to-integrate2}
\end{multline}
These terms will integrate to zero given some special symmetries.

The natural symmetry to consider is time-reversal symmetry.
Indeed, this helps, but only partially, since for a time-reversal preserving system, we have
\begin{align}
  T^{-1} h(\mathbf -p) T & = h(p) \\
  \mathbf d(-\mathbf p) \cdot T^{-1} \bm \sigma T & = \mathbf d(\mathbf p) \cdot \bm \sigma.
\end{align}
As an anti-unitary operator, $T = U K$  where $U$ is a unitary and $K$ is complex conjugation.
Complex conjugation will just let $\sigma_y \rightarrow -\sigma_y$, and $U$ will rotate along the Bloch sphere.
Thus, any sort of triple product will change sign, and we have 
\begin{align}
  \partial_x d_0 [\partial_y \hat{\mathbf d} \cdot (\dd_0 \times \dd )]|_{\mathbf p \rightarrow -\mathbf p} = -\partial_x d_0 [\partial_y \hat{\mathbf d} \cdot(\dd_0 \times \dd) ].
\end{align}
Naturally $d_0(-p) = d_0(p)$ as well.
Thus, with both momenta connected by time-reversal symmetry, they lie at the same energy and any integral over a finite chemical potential will cancel their contributions.

However, there is still the possibility that the term that goes as $\cos 2d_0 t_1$ in Eq.~\eqref{eq:persistentHall-to-integrate} will be finite.

To address this term, if both systems respect a mirror symmetry (unitary or anti-unitary) along one plane, this term will also vanish.
In particular, let us say the ``mirror plane'' is along the $x$-direction without loss of generality.
Then, define the mirror symmetry operator $M_y = U K^a$ (where $a=0$ if unitary and $a=1$ if anti-unitary) as acting on $H$ such that
\begin{align}
  M^{-1}_y h(p_x,-p_y) M_y &  = h(p_x, p_y)  \\
  \tilde I\mathbf d(p_x, -p_y) \cdot \bm\sigma & = \mathbf d(p_x,p_y) \cdot \bm \sigma,
\end{align}
where $\tilde I$ is either the identity or an inversion operator on $\mathbf d$. In fact, $\tilde I$ should act as the identity on $d(p_x,0)$, and so all $d(p_x,0)$ span an invariant subspace for $\tilde I$. Generally this space will be more than 1 dimensional, and so if it is two-dimensional (if it is three-dimensional, $\tilde I$ will just be the identity), we have
\begin{align}
   h(p_x,0) = d_x(p_x,0) \sigma_x + d_z(p_x,0) \sigma_z,
\end{align} 
where without loss of generality we chose the $x$- and $z$-directions to be the invariant directions. Thus, we have $d_y(p_x,p_y)\rightarrow 0$ as $p_y\rightarrow 0$. Furthermore, inversion can only occur in that direction, so we have
\begin{align}
  \mathbf d(p_x,-p_y) = (d_x( p_x,p_y), \pm d_y(p_x,p_y), d_z(p_x,p_y)).
\end{align}
This allows us to say
\begin{align}
  \partial_x d_0 \partial_y \hat{\mathbf d}_0 \cdot \dd |_{p_y \rightarrow -p_y} = - \partial_x d_0 \partial_y \hat{\mathbf d}_0 \cdot \dd.
\end{align}
And by similar reasoning as before, once we integrate over all momenta, there will be no contribution to the Hall current.

Thus, we have proved the following statement:
\emph{If $h(\mathbf p, \Delta)$ and $h(\mathbf p, 0)$ both have time-reversal symmetry and mirror-symmetry along the same plane, then $\Sigma_{\rm Hall}^\infty = 0$.}

Another way to phrase this is: if $h(\mathbf p, \Delta)$ has mirror symmetry along some axis for all $\Delta$, then time-reversal symmetry breaking before and after the quench implies $\Sigma_{\rm Hall}^\infty = 0$.
The models we study are just that: Models that have such a mirror symmetry.

\subsubsection{Relation to Berry curvature}

We can now make a more precise claim about the Hall current's relation to Berry curvature.

First, we take the equation for Larmor precession Eq.~\eqref{eq:LarmorPrecession1} and rewrite it as
\begin{align}
  \hat{\mathbf d}_0  = \frac{\hat{\mathbf n} \times \partial_t \hat{\mathbf n}}{2d_0}- \hat{\mathbf n}(\hat{\mathbf d}_0 \cdot \hat{\mathbf d}).
\end{align}
The term that is important for Eq.~\eqref{eq:persistentHall-to-integrate} is
\begin{align}
  \hat{\mathbf n} \cdot \partial_\mu \hat{\mathbf d_0} = \hat{\mathbf n} \cdot \left[\frac{\partial_\mu\hat{ \mathbf n} \times \partial_t \hat{\mathbf n}}{2d_0} \right] - \partial_\mu(\hat{\mathbf d}_0 \cdot \hat{\mathbf d}).
\end{align}
Now, we can write the Berry curvature as $\Omega_{\mu t} = \frac12 \hat{\mathbf n} \cdot (\partial_\mu\hat{ \mathbf n} \times \partial_t \hat{\mathbf n})$, and thus, we have
\begin{multline}
  \chi^\infty_{\rm Hall}(\mathbf p, t_1) = e^2 \{\partial_y \log d_0 [\Omega_{x t_1}-\partial_x(\hat{\mathbf d}_0 \cdot \hat{\mathbf d})] \\- \partial_x \log d_0 [\Omega_{y t_1}-\partial_y(\hat{\mathbf d}_0 \cdot \hat{\mathbf d})]\}.
\end{multline}
If we assume the system has the mirror-symmetry described in the preceding section, then
\begin{align}
  \Sigma_{\rm Hall}^\infty = e^2 \int_p [\Omega_{x t_1} \partial_y \log d_0 - \Omega_{y t_1} \partial_x \log d_0 ].
\end{align}
This simplifies one step further if we began in a filled band, then we can integrate by parts without picking up boundary terms and we get
\begin{align}
  \Sigma_{\rm Hall}^\infty =  e^2 \int_p \partial_{t_1} \Omega_{yx} \log d_0.
\end{align}
The response of the system is purely described in terms of a weighted integral over the \emph{derivative} of the Berry curvature. This Berry curvature is evaluated for the state at the time of the pulse.

\subsubsection{Full time current response}

For completeness, we just mention that if we define $\hat{\mathbf n}_2 = \braket{\psi_2 | \bm \sigma| \psi_2}$, we can easily write the expression for the single particle current response:
\begin{align}
 j_\mu(\mathbf p; t_2, t_1) = j_\mu^\infty(\mathbf p; t_1) + \tfrac{e}2 \partial_\mu \hat{\mathbf d}_A \cdot (\hat{\mathbf d}_A \times \partial_{t_2} \hat{\mathbf n}_2),
\end{align}
where we just defined $\hat{\mathbf d}_A = \hat{\mathbf d}(\mathbf p - e \mathbf A, 0)$.

\section{The time-reversal broken BHZ model}

Lastly, we consider the time-reversal broken BHZ model with
\begin{align}
  h(\mathbf k, M) = \mathbf d(\mathbf k, M)\cdot \bm \sigma,
\end{align}
where
\begin{align}
  \mathbf d(\mathbf k, M) = (\sin k_x, \sin k_y, M + 2 - \cos k_x - \cos k_y).
\end{align}
This model admits 4 gapped (or ungapped at the phase transitions) Dirac cones. 
In equilibrium, depending on the sign of the gap at each point (controlled by $M$), we can have a positive, negative or zero Hall conductance overall.

Generically, we will quench from $M$ to $M'$ and use Eq.~\eqref{eq:persistent-current-gen-expr} to determine the long-time persistent response.
The ingredients we need are (1) the ground state of $h(\mathbf k, M)$, (2) the time evolution with the quenched Hamiltonian $h(\mathbf k, M')$, and (3) the eigenstates and energies of $h(\mathbf k - e\mathbf A, M')$.

We shall denote our resulting $d$-vectors by $h(\mathbf k, M) = \mathbf d \cdot \bm \sigma$, $h(\mathbf k, M') = \mathbf d_0 \cdot \bm \sigma$, and $h(\mathbf k - e \mathbf A, M') = \mathbf d_A \cdot \bm \sigma$.

We can then write the ground state of $h(\mathbf k, M)$ as
\begin{align}
  \ket{\psi_0} = \frac{(d - d_z ) \ket\uparrow - (d_x + i d_y) \ket{\downarrow}}{\sqrt{2d(d- d_z)}},
\end{align}
the time evolution of $h_0 = h(\mathbf k, M')$ as
\begin{align}
  e^{-i t h_0} = \begin{pmatrix}
    \cos t d_0 - i\frac{d_{0,z}}{d_0} \sin t d_0 & -i\frac{d_{0,x} - i d_{0,y}}{d_0} \sin t d_0 \\
    -i\frac{d_{0,x} + i d_{0,y}}{d_0} \sin t d_0 & \cos t d_0 + i\frac{d_{0,z}}{d_0} \sin t d_0
  \end{pmatrix},
\end{align}
and the energies and eigenstates of $h_A = h(\mathbf k - e\mathbf A, M')$ as
\begin{align}
   \epsilon_{A,\pm} & = \pm d_A, \\
   \ket{\epsilon_{A,\pm}} & = \frac{(d_A \pm d_{A,z} ) \ket\uparrow \pm (d_{A,x} + i d_{A,y}) \ket{\downarrow}}{\sqrt{2d_A(d_A \pm d_{A,z})}}.
\end{align}

With these ingredients we can solve for the persistent current for a single momentum $\mathbf k$, and then integrate over the Brillouin zone to obtain the total persistent current.

\begin{figure*}
\includegraphics[width=2\columnwidth]{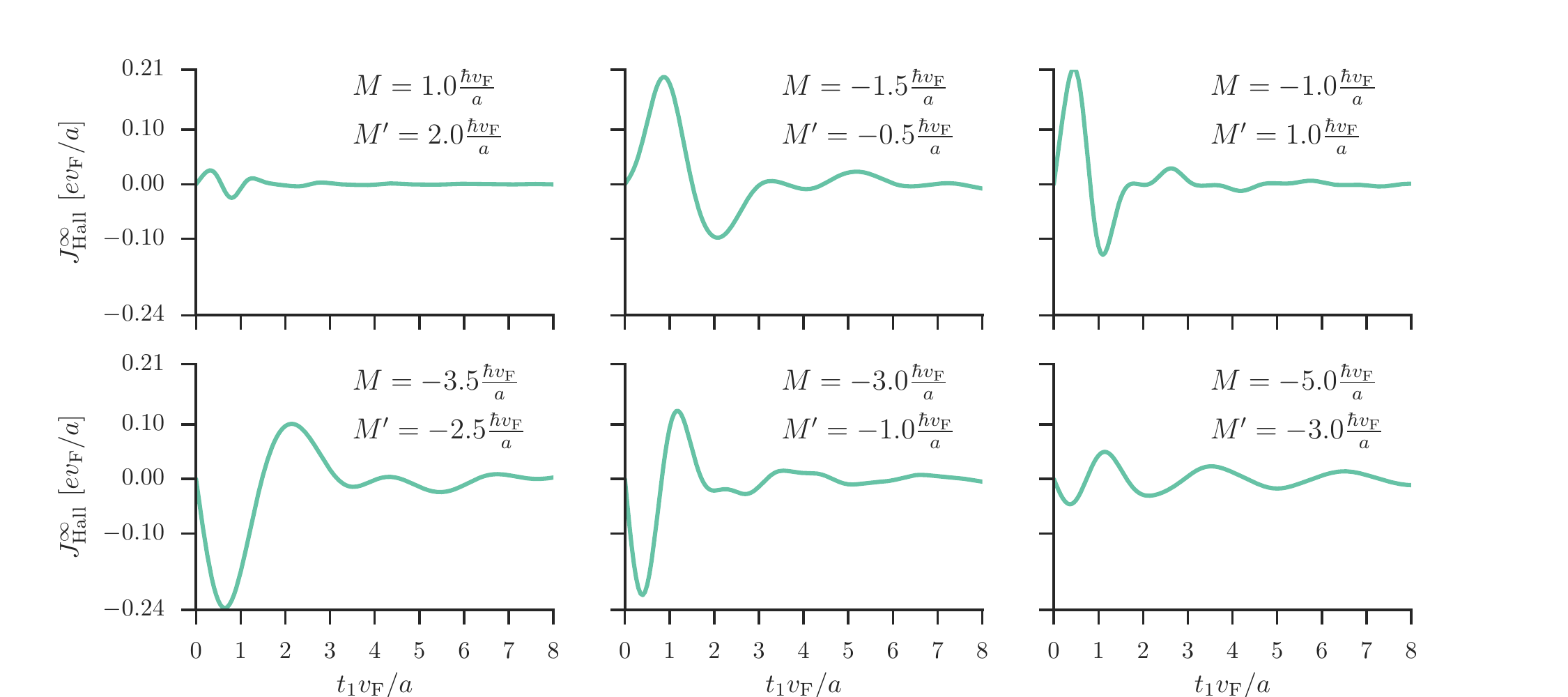}
\caption{The long-time ballistic current versus the time of the pulse $t_1$ for the time-reversal-broken BHZ model. Notice how that depending on when the pulse hits, the ballistic current can be positive or negative. Additionally, the finite size keeps there from being a response when $t_1\rightarrow 0^+$, and there a `turn-on' time for the ballistic current (a time we must wait to get the largest effect). The pulse used in these figures is $A_x = 0$ and $e A_y = 0.1 \hbar/a$.}
\label{fig:BHZmodel}
\end{figure*}

This procedure can be done easily numerically, and we obtain what is plotted in Fig.~\ref{fig:BHZmodel} (we restore physical units in the plot).
Notice that we get an effect independent of what phase we are quenching from or to. 
For reference, the phases are trivial for $M>0$, $-2<M<0$ is a topological insulator with $\sigma_{xy} = -1$, $-4<M<-2$ is a topological insulator with $\sigma_{xy}=+1$, and $M<-4$ is back to a trivial insulator.

\section{Analogy with monopoles and dipoles}

In these models with nontrivial Berry curvature, we can say that we have monopoles that are sources of Berry curvature.
Naturally extending this, in the Haldane model represented by $h = v_\mathrm{F} \mathbf k \cdot \sigma + \Delta \sigma_z$, the (positively charged) monopole has been lifted above the $x$-$y$ plane.

The quenching procedure is then suddenly moving this charge close to the plane, and the pulse would be a sudden movement parallel to the plane. 
Meanwhile the dipoles, having not relaxed, are rotating around their local magnetic field -- Larmor precessing.
The statement above then becomes that a net magnetization will appear perpendicular to the last movement.
However, the precession is quite strange: The magnetic field strength (and hence how fast the precession happens) is increased for states further away from the monopole.

This will give us a slightly different effect, and we can easily approach this problem with some of the geometric machinery we used before.
Assume that there is a monopole at a position $z_0$ above a plane of non-interacting dipoles, then the dipoles will align themselves with the monopole such that at position $\mathbf r = (x,y)$
\begin{align}
  \mathbf M = \mu \frac{\mathbf r - z_0 \hat{\mathbf z} }{|\mathbf r - z_0 \hat{\mathbf z}|}.
\end{align}
The charge from the monopole is spherically symmetric, but we will not specify how it depends on the distance yet
\begin{align}
  \mathbf B(\mathbf r) = B(r) \hat{\mathbf r}.
\end{align}
The monopole begins with $\mathbf B(\mathbf r - z_0 \hat{\mathbf z})$.
Then, we suddenly shift the monopole down in plane.
Each dipole feels magnetic field from the monopole and rotates as described by the equation of motion
\begin{align}
  \frac{d \mathbf M}{d t} = \mathbf M \times \mathbf B,
\end{align}
In fact, we can write
\begin{align}
  \mathbf M = - \mu \frac{\mathbf r + z_0 [\cos B(r) t \, \hat{\mathbf z} + \sin B(r) t \, (\hat{\mathbf z} \times \hat{\mathbf r})]}{\sqrt{r^2 + z_0^2}}.
\end{align}
Then at $t=t_1$, we suddenly shift the monopole in the $x$-direction by an amount $x_1$, and we use the formula to determine how much the center of rotation is along the $y$-axis:
\begin{align}
  \overline{M}_y & = \frac{[\mathbf M(t_1) \cdot \mathbf B(\mathbf r - x_1 \hat{\mathbf{x}})][\mathbf B(\mathbf r - x_1 \hat{\mathbf{x}}) \cdot \hat{\mathbf y}]}{B(\mathbf r - x_1 \hat{\mathbf{x}})^2}, \\
    & = \frac{[\mathbf M(t_1) \cdot (\mathbf r - x_1 \hat{\mathbf{x}})][(\mathbf r - x_1 \hat{\mathbf{x}}) \cdot \hat{\mathbf y}]}{(\mathbf r - x_1 \hat{\mathbf{x}})^2}
\end{align}
Evaluating this
\begin{align}
  \overline{M}_y & = -\mu \frac{y}{\sqrt{r^2 + z_0^2}}\frac{r^2 - x_1 x + z_0 x_1 \frac{y}{r} \sin B(r) t_1}{r^2 - 2x_1 x + x_1^2}.
\end{align}
Now, assume the dipoles have a density $\rho$ on the $xy$-plane, when we integrate this function over the plane, we can break it into polar coordinates, but notice that the only nonzero term will be the term which goes as $\sin B(r) t_1$ above (everything else is odd in terms of $y$ and can be safely discarded).

Therefore, our total magnetization that dynamically appears perpendicular to the shift in the monopole, $M^\infty_y$, can be written as
\begin{multline}
  M^\infty_y = -\mu \rho z_0 x_1 \int_0^\infty dr \int_0^{2\pi} d\phi \frac{r^2 \sin^2\phi}{r^2 - 2x_1 r \cos \phi + x_1^2} \\ \times \frac{\sin B(r) t_1}{\sqrt{r^2 + z_0^2}}.
\end{multline}
Performing the angular integral,
\begin{multline}
M^\infty_y = -\pi \mu \rho z_0 x_1 \left[ \int_{0}^{x_1} dr\left(\frac{r}{x_1}\right)^2 \frac{\sin B(r) t_1}{\sqrt{r^2 + z_0^2}}\right. \\  \left. + \int_{x_1}^{\infty} dr \frac{\sin B(r) t_1}{\sqrt{r^2 + z_0^2}} \right].
\end{multline}
If we assume that $x_1$ is small, we can disregard the first integral as a small correction, and we can approximate the last integral by
\begin{align}
  \int_{x_1}^{\infty} dr \frac{\sin B(r) t_1}{\sqrt{r^2 + z_0^2}}  \sim \int_{0}^{\infty} dr \frac{\sin B(r) t_1}{\sqrt{r^2 + z_0^2}}. 
\end{align}
If we further now assume that $B(r) = g/r^2$, then we have 
\newcommand{\argum}{\alpha}
\begin{align} 
  \int_0^\infty dr  \frac{\sin (g t_1/r^2)}{\sqrt{r^2 + z_0^2}} & = 2\int_0^\infty \frac{dw}{w} \frac{\sin w}{\sqrt{1+ \frac{z_0^2}{g t_1} w }} \\
   & = 2\pi \left\{ C\left(\argum\right)\left[ 1 - C\left(\argum\right) \right] \right. \nonumber \\ & \phantom{ = \pi \{ C(a\} } \left. + S\left(\argum\right)\left[ 1 - S\left(\argum\right) \right] \right\},
\end{align}
where $\alpha = \sqrt{\tfrac2\pi}\tfrac{g t_1}{z_0^2}$ and $S$ and $C$ are the Fresnel $S$ and $C$ integrals.
Just as before, we can expand for short and long $t_1$ times.
The results are
\begin{align}
  M_y^\infty & \sim -\pi^2 \mu \rho z_0 x_1,  & \tfrac{g t_1}{z_0^2} \gg 1, \quad x_1\ll 1, \\
  M_y^\infty & \sim -\pi \mu \rho x_1 \sqrt{2\pi g t_1}, & \tfrac{g t_1}{z_0^2} \ll 1, \quad x_1 \ll 1.
\end{align}
Thus, we see that the $1/r^2$ force yields slightly different results (notably, that the result for long $t_1$ times is a constant), but the overall idea and analogy are much the same.

\end{document}